\documentclass{PoS}
\DeclareGraphicsExtensions{.jpg,.pdf,.png,.eps,.pdf}
\graphicspath{ {pictures/} }

\usepackage{amssymb}
\usepackage{graphicx}
\usepackage{url}
\usepackage{color}
\usepackage{tabulary}
\usepackage{multirow}
\usepackage{amsmath}
\usepackage{amsfonts}
\usepackage{ulem}
\newcommand{\simlt}{\mathrel{\hbox{\rlap{\hbox{\lower4pt\hbox{$\sim$}}}\hbox{$<$}}}}
\newcommand{\simgt}{\mathrel{\hbox{\rlap{\hbox{\lower4pt\hbox{$\sim$}}}\hbox{$>$}}}}

\newcommand{\erg}{\;\mathrm{erg}}
\newcommand{\s}{\;\mathrm{s}}
\newcommand{\dd}{\partial}
\newcommand{\Msol}{\;\mathrm{M}_{\odot}}
\newcommand{\cm}{\;\mathrm{cm}}
\newcommand{\AU}{\;\mathrm{AU}}

\newcommand{\yr}{\;\mathrm{yr}}
\newcommand{\Myr}{\;\mathrm{Myr}}
\newcommand{\keV}{\;\mathrm{keV}}

\def\apj{\textit{ApJ~}}
\def\mnras{\textit{MNRAS~}}
\def\aap{\textit{A\&A~}}

\def\araa{\textit{Ann. Rev. A\&A~}}

\def\apjs{\textit{ApJ Supp.~}}

\title{Population III X-Ray Binaries}

\ShortTitle{Population III X-Ray Binaries}

\author{\speaker{Taeho Ryu}, Takamitsu L. Tanaka and Rosalba Perna\\%
	The State University of New York, Stony Brook, USA\\
	E-mail: \email{taeho.ryu@stonybrook.edu}}

\abstract{Understanding of the role of X-rays for driving the thermal evolution of the intergalactic medium (IGM) at high redshifts 
	is one of important questions in Astrophysics. High-mass X-ray binaries (HMXBs) in early stellar populations
	are prime X-ray source;
	however, their formation efficiency is not well understood.
	Using $N$-body simulations, we estimate the HMXB formation rate
	via mutual gravitational interactions of nascent, small groups of the Population~III stars. 
	We find that HMXBs form at a rate of one per $\simgt 10^{4}\Msol$ in newly born stars,
	and that they emit with a power of $\sim 10^{41} \erg \s^{-1}$ in the $2-10$ keV band 
	per star formation rate (SFR). 
	This value is a factor $\sim 10^{2}$
	larger than what is observed in star forming galaxies at lower redshifts;
	the X-ray production from early HMXBs would have been even more copious,
	if they also formed \textit{in situ} or via migration in protostellar disks.
	Combining our results with earlier studies suggests that early HMXBs
	were highly effective at heating the IGM and leaving a strong 21 cm signature.
	We discuss broader implications of our results,
	such as the rate of long gamma-ray bursts from Population~III stars
	and the direct collapse channel for massive black hole formation.}

\FullConference{Frontier Research in Astrophysics  II\\
	23-28 May 2016\\
	Mondello (Palermo), Italy}

\begin{document}
	
	\section{Introduction}
	\label{sec:intro}

	One of the fundamental challenges in cosmology is to understand the history
	of the early Universe from Cosmic dark ages to the end of reionization as the first stars and galaxies emerged and the radiation from these objects had ionized the intergalactic medium (IGM).
	Advances in numerical techniques,
	combined with exquisite measurements of the ``initial'' conditions (\cite{Hinshaw+13}, \cite{Planck15}),
	have led to remarkable simulations (\cite{Abel+02}, \cite{Turk+09},  \cite{Stacy+10}, \cite{greif11}, \cite{BrommYoshida11}) of the conditions leading
	up to the former milestone, occurring at $z\simgt 30$, when the Universe was $\approx 100\Myr$ old.
	However, reconstructing the subsequent several hundred Myr of cosmic history has embraced difficulties, one of which 
	is related to uncertainties in modeling the numerous forms of feedback from
	the first astrophysical objects (\cite{Springel+05}, \cite{Stinson+06}, \cite{Sijacki+07}).
	
	In particular, X-rays from the first galaxies can act as a powerful
	source of feedback (\cite{Venkatesan+01}, \cite{Machacek+03}). Because hard
	X-rays (energies $\simgt 1\keV$) have mean free paths comparable to
	the Hubble horizon, they can isotropically heat and partially reionize
	the early IGM (\cite{Peng2001}, \cite{Venkatesan+03}, \cite{Ricotti2004},
		\cite{Pritchard2007}). Recent studies show that they are expected to be the dominant
	agent in heating the IGM, possibly leading to suppression of star formation
	(\cite{Ripamonti+08}) and massive black hole (BH) growth
	(\cite{Tanaka2012}) inside low-mass dark matter haloes by raising the
	Jeans and filtering masses of the IGM (\cite{Gnedin00}, \cite{NaozBarkana07}).
	On larger scales, as soft X-rays ($\sim 0.1-1\keV$) promotes the formation
	of molecular hydrogen, affecting the formation of stars and possibly massive black holes
	(BHs) (\cite{Haiman+96}, \cite{Kuhlen2005},
		\cite{Latif+15}, \cite{Inayoshi2015}).

	The goal of this study is to estimate the formation rate and X-ray output
	of HMXBs in the early Universe as one of the prime X-ray sources, by using $N$-body simulations of
	nascent groups of the first (Population III, henceforth Pop~III) stars. 
	The formation rate has been left as a free parameters in studies or 
	estimated with simplified assumptions. However, we estimate this 
	value via purely theoretical dynamical approach.
	Based on the estimated formation rates, we derive a HMXB energy output (normalized to the star formation rate)
	that is a factor $\sim 10 - 150$ higher than in present-day star-forming galaxies. We find that the X-ray output does not change significantly within the wide variety
	of simulation setups and submit that this is a robust estimate. Our work is also relevant for predicting the rates of
	long-duration gamma-ray bursts (LGRBs) from Pop~III stars.

	The paper is organized as follows. We start in \S\ref{sec:dynamics}
	by discussing the problem to be solved---beginning with the equations
	of motion, followed by the description of our $N$-body code,
	our choices for the initial conditions,
	and how the data is interpreted for HMXB formation.
	We present our results in \S\ref{sec:results}.
	In \S\ref{sec:discussion}, we discuss the implication of our
	work for the X-ray output of the first galaxies, as well
	as for other topics such as LGRBs and SMBH formation.
	We conclude with a summary of our findings in \S\ref{sec:summary}.

	\section{Stellar Dynamics}\label{sec:dynamics}
	
	Here, we provide an overview of our simulations---namely:
	the equations of motion that are solved to
	simulate the dynamical evolution of the star groups;
	the numerical scheme we use to solve the equations;
	the different types of initial conditions we adopted,
	as well as the reasoning behind our choices;
	and finally, how the results are interpreted for HMXB formation.
	
	\subsection{The equations of motion}
	\label{subsec:theequationofmotion}
	
	Our $N$-body code numerically solve the equation of the motion of $N$ objects with masses
	$m_{i}$, moving under their mutual gravitational influence,
	a dissipative dynamical friction force,
	and a background gravitational potential, which can be written as, 
	\begin{equation}\label{eq:eom}
		\frac{d^{2}}{dt^{2}}\vec{r}_{i}=
		\vec{a}_{{\rm g}, i}+\vec{a}_{{\rm df}, i}+\vec{a}_{{\rm bg}, i}.
	\end{equation}

	The first term on the right-hand side of equation (\ref{eq:eom}) is the specific force due to
	Newtonian gravity,
	\begin{equation}
		\vec{a}_{{\rm g}, i}
		= -\sum_{j\neq i}
		G~m_{j}~
		\frac {\dd ~S(r_{ij})}{\dd ~r_{ij}}~
		\frac{\vec{r}_{i}-\vec{r}_{j}}{r_{ij}},
	\end{equation}
	where $G$ is the gravitational constant, 
	$\vec{r}_{i}$ is the displacement of the $i_{{\rm th}}$ star
	from the center of the host dark matter halo,
	and $r_{ij}\equiv |\vec{r}_{i}-\vec{r}_{j}|$.
	
	We adopt the Plummer softening kernel $S(r_{ij})$ (\cite{Galacticdynamics}),
	
	\begin{equation}\label{eq23}
		S(r_{ij})=-\frac{1}{\sqrt{r_{ij}^{2}+\epsilon^{2}}}\,,
	\end{equation}
	where we take $\epsilon=R_{\odot}$.
	
	The second term on the right-hand side of equation (\ref{eq:eom}), $\vec{a}_{{\rm df}, i}$,
	is the specific drag force due to dynamical friction.
	For collisionless systems, the standard Chandrasekhar formula for
	dynamical friction is \cite{Galacticdynamics},
	\begin{equation}\label{eq:df1}
		\vec{a}_{i}=-4\pi ~\ln\Lambda ~f(X_{i}) ~ \frac{G^{2} m_{i}}{v_{i}^{3}}~\rho(r_{i})~\vec{v}_{i},
	\end{equation}
	where 
	\begin{equation}\label{eq:df2}
		f(X_{i})\equiv {\rm erf}(X_{i})-\frac{2}{\sqrt{\pi}} ~X_{i} ~\exp\left(-X_{i}^{2}\right),
	\end{equation}
	
	$v_{i}$ is the speed of the $i_{th}$ star with respect to the background,
	$X_{i}\equiv v_{i}/(\sqrt{2} \sigma_{v})$,
	$\sigma$ is the velocity dispersion,
	$\ln\Lambda$ is the Coulomb logarithm and
	$\rho(\vec{r}_{i})$ is the local gas density.
	
	We adopt the modified formula for gaseous medium used in \cite{dyformula1}.
	This prescription incorporates behaviors found in numerical simulations for
	subsonic and supersonic regimes (\cite{dyformula2}, \cite{dyformula3}).
	The specific drag force vector always points opposite to the direction
	of motion. We use $\ln\Lambda=3.1$.

	The third and last term, $\vec{a}_{{\rm bg}, i}$,  is the specific force due to
	the background potential, which is dominated by gas. 
	The background potential provides an
	additional inward force whose functional form depends on the density profile.
	For simplicity, here we use a constant density and explore different values in our simulations.
	The force due to the background potential is then
	\begin{equation}
		\label{eq17}
		\vec{a}_{{\rm bg}, i}=-\frac{4 \pi}{3}G\rho \vec{r}_{i},
	\end{equation}
	where here $\vec{r}_{i}$ is the vector pointing from the 
	center of the halo to the $i$-th star.

	The equation of motion is solved iteratively, with
	the positions, velocities and accelerations
	of each star updated at every time step.
	We describe our computational method below.
	
	\subsection{Code Description}
	\label{subsec:codedescription}
	
	We perform 3-dimensional, $N$-body simulations with 4th-order \& 5-stage
	Runge-Kutta-Fehlberg methods (RKF45 method, \cite{Fehlberg}) using
	adaptive time steps. The RKF45 is a very precise and stable
	integration method among the large class of Runge-Kutta schemes,
	particularly by adapting the Butcher tableau for Fehlberg's 4(5)
	method.
	
	The code keep updating the position and velocity by integrating the equations of motion (\ref{eq:eom}), taking into account the effects of dynamical frictions and gravitational force from ambient gas.
	
	To ensure numerical precision, our computational scheme varies the
	value of each subsequent time step analytically, so that
	numerical errors for each variable in the simulation do
	not exceed $10^{-13}$ times the size of the variable. 
	
	\subsection{Determining HMXB formation}
	\label{subsubsec:DeterminingHMXB}
	It is assumed that a HMXB has formed if 
	both of the following criteria are satisfied:
	\begin{enumerate}
		\item \textit{One of two stars forming a binary turns into
			a compact object (CO).} We compare the typical lifetime
		($\tau_{{\rm life}}$) of a massive star with the time ($t_{{\rm run}}$)
		in the simulation (taken to coincide with the time at which stars are
		born). If $\tau_{{\rm life}}>t_{\rm run}$, the star is marked as a
		CO in the simulation. 
		
		\item \textit{The two stars are close enough so that the accretion occurs
			through Roche-lobe overflow (RLOF)}.
		This criterion is simply written as $R_{{\rm star}}\geq
		R_{{\rm RL}}$, where $R_{{\rm star}}$ is the stellar radius, and 
		$R_{{\rm RL}}$ is the Roche-lobe radius of the most massive star calculated from the
		center of the star to the inner Lagrange point given by \cite{Rocheradius},
		\begin{equation}\label{eq7}
			\frac{R_{{\rm RL},1}}{r}=\frac{0.49q^{2/3}}{0.6q^{2/3}+\ln(1+q^{1/3})}\,,
		\end{equation}
		where $r$ is the orbital distance and $q=\frac{m_{2}}{m_{1}}$ is the
		mass ratio. 
		
	\end{enumerate}
	
	If a binary satisfies these two criteria, the binary is marked as a
	HMXB. Since the two criteria are independently checked
	at every time step,
	which criterion is satisfied first is not important.

	\subsection{Setup and Initial Conditions}
	\label{sec:initialsetup}
	
	We model dark matter halo with virial temperatures $\sim 1000-2000\;{\rm K}$ at the redshift $20\simlt z\simlt 30$ as a birth place of Pop~III formation.
	The initial positions of the stars are generated quasi-randomly via a Monte Carlo
	realization with the assumption that they formed inside a shared Keplerian gas disk at the center of the host dark matter halo. Correspondingly, the initial velocities of the stars are assigned to be circular, parallel to the disk plane, and Keplerian at the instant the simulation begins.
	
	 In the simulations by
		\cite{Greif12}, multiple protostars formed several AU apart from each other.
		The masses of the stars follow the initial mass function (IMF)
		with $\alpha=0.17$ ($\frac{dN}{dM}=M^{-\alpha}$, \cite{stacy13}),
				$M_{\rm max}=140\Msol$ and
				$M_{\rm min} =0.1\Msol$. 
		we run simulations with number densities  $n_{6}$ and $n_{4}$. 
		In addition to that, we explore several different configurations:
		cases with a star group of $N=5$ stars,
		a star group of $N=10$ stars,
		and collisions between two groups of $N=5$ stars.

	\section{Results}
	
	Here, we summarize the findings of our $N$-body simulations, 
	focusing in particular on the properties of the most compact binaries
	found for each set of runs.
	
	\label{sec:results}

	\subsection{5-body simulations}
	\label{subsec:smallscale}
	
	\begin{table}
		\centering
		\setlength\extrarowheight{4pt}
		\begin{tabulary}{0.7\linewidth}{c c c}
			\hline
			  & $n_{6}$ & $n_{4}$\\
			\hline
			runs &			86					&	86\\
			$\langle a_{t=500{\rm yr}}\rangle$ [AU]   &  	1.37					&  	1.42						  \\
			$\tau_{{\rm df}}$[yr]	 &     $\sim10^{13}$		 &   $\sim 10^{15}$			  \\
			Companion stars 		&	\multicolumn{2}{c}{3rd massive star, $11\sim12 \Msol$}\\
			$P_{{\rm HMXBc}}$ &    0.070	 &  0.070			  \\
			$F_{\rm HMXB}$ [$10^{-4}\Msol^{-1}$] &    4.6	 &  4.6		\\
			\hline
		\end{tabulary}
		\caption{Summary of results for simulations of 5-body
			groups forming on scales of $\simeq 10\AU$.
			$\langle a_{t=5000~{\rm yr}}\rangle$
			indicates the semi-major axis at $t=5000$ yr, while
			$\tau_{{\rm df}}$ represents the dynamical friction
			timescale required for $a_{t=5000~{\rm yr}}$ to
			shrink to $a_{{\rm RL}}$.
			$P_{\rm HMXBc}$ is the fraction of runs in which a  HMXBc forms,
			and $F_{\rm HMXB}$ is the number of HMXBc formed across
			all simulations, normalized by the total mass of the stars in the simulations.                  
		}
		\label{tab:tab2}
	\end{table}

	\begin{figure*}
		\centering
		{\includegraphics[width=7.5cm]{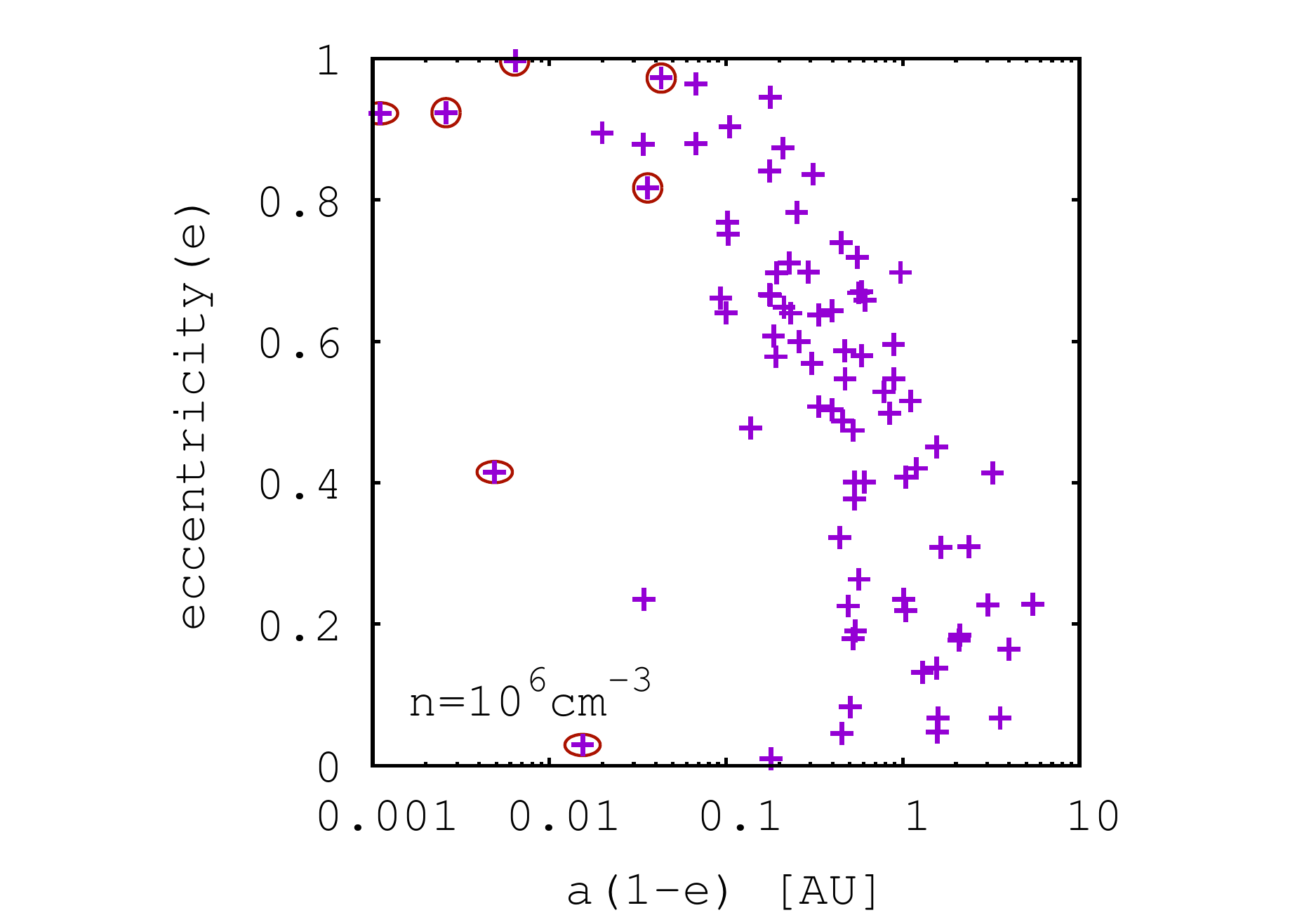}}
		{\includegraphics[width=7.5cm]{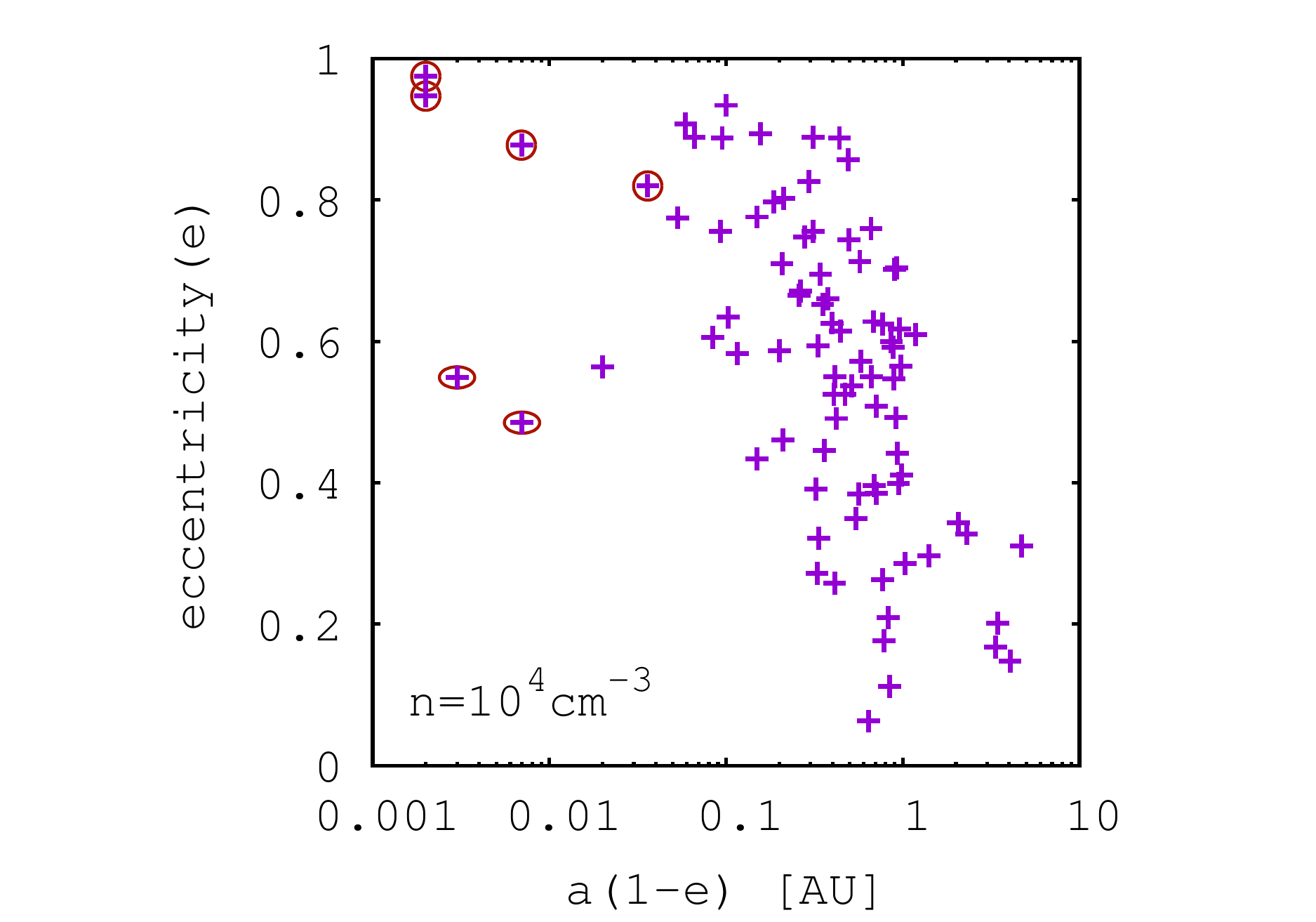}}
		\caption{ {\it Left}: Pericenter distance - eccentricity
			distribution plot for $n_{6}$. Two HMXB candidates (HMXBc)
			have been produced. {\it Right}: The same distribution plot
			for $n_{4}$. The circled points indicate HMXBc whose 
			mass transfer via RLOF may occur periodically due to their
			eccentric orbits. The elliptically-circled point indicates an
			HMXBc whose semimajor axis is smaller than $a_{{\rm RL}}$, so
			the mass transfer will be steady. There are a
			couple of binaries with pericenter distance smaller than
			$R_{{\rm RL}}$, but they are excluded because $M_{2}< 8 \Msol$.}
		\label{fig:8}
	\end{figure*}

	We performed 86
	runs of 5-body simulations for each of the number density values $n_6$ and $n_4$. We use the same initial conditions for
	each set of runs for proper comparisons.

	In the $n_{6}$ calculations, the most common scenario
	is that ${\rm S}_{1}$ always forms the most compact binary almost immediately,
	while stellar scatterings are most common during 
	the first few years to about 40 years.
	Thereafter, a multiple system usually survives and stabilizes, while less massive stars are ejected.

	In our 86 runs, the last surviving binary is typically of type ${\rm
		B}_{13}$ (pairing of the most massive and third most massive stars).
	with total mass of $130\Msol$ and $\langle M_{{\rm
			S}_{3}}\rangle=11\Msol$. $\langle a_{t=5000~{\rm
			yr}}\rangle=1.87\AU$ and the minimum semi-major axis is
	$0.2\AU$. The corresponding dynamical friction time scale is $\sim 10^{13}$ yr. The ejected stars have speeds of
	$(10-100)~c_{\rm s}$ , and $c_{\rm s}\sim 4~{\rm km/s}$.
	Six of the HMXB candidates (HMXBc hereafter) have been formed in all
	runs ($P_{{\rm HMXBc}}$ = 0.070). For later use, let us define
	$F_{\rm HMXB}$ as the number of HMXBc formed across all simulations,
	normalized by the total mass of the stars. Then $F_{{\rm HMXBc}} =
	4.6\times 10^{-4} \Msol^{-1}$ (This term will be used to estimate
	the X-ray luminosity and is one of the primary results of our
	study).
	
	The outcomes of the $n_4$ simulations are quite similar: a triple or
	higher multiple forms in 65 of 86 runs, and $\langle
	a_{t=5000~{\rm yr}}\rangle=1.42\AU$. Furthermore, due
	to the same number of HMXBc, $P_{{\rm HMXBc}}$ and $F_{{\rm HMXBc}}$
	are the same.  One notable difference is that for $n_4$, the
	dynamical friction time scale is longer by 2 orders of magnitude
	compared to $n_{6}$, because this quantity is inversely proportional
	to the number density.
	
	We present the distribution of eccentricity for
	pericenter distance in Figure~\ref{fig:8}. In particular, the {\it left}
	panel shows the distribution for $n_{6}$ calculations and the {\it
		right} panel for $n_{4}$. The circled point indicates HMXBc with
	distance at pericenter shorter than the corresponding Roche-Lobe
	radius.

	\begin{figure*}
		\centering
		{\includegraphics[width=7.5cm]{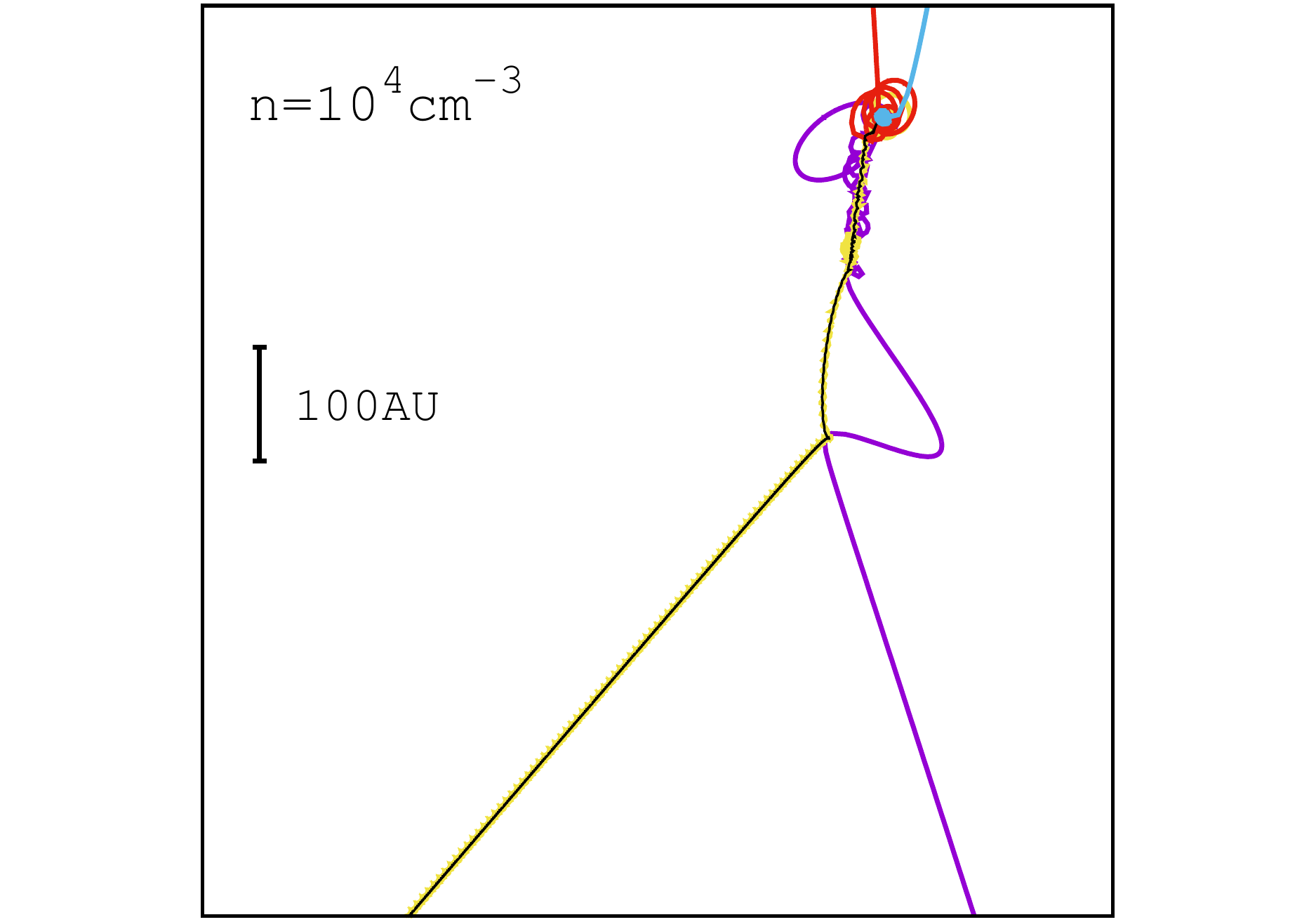}}
		{\includegraphics[width=7.5cm]{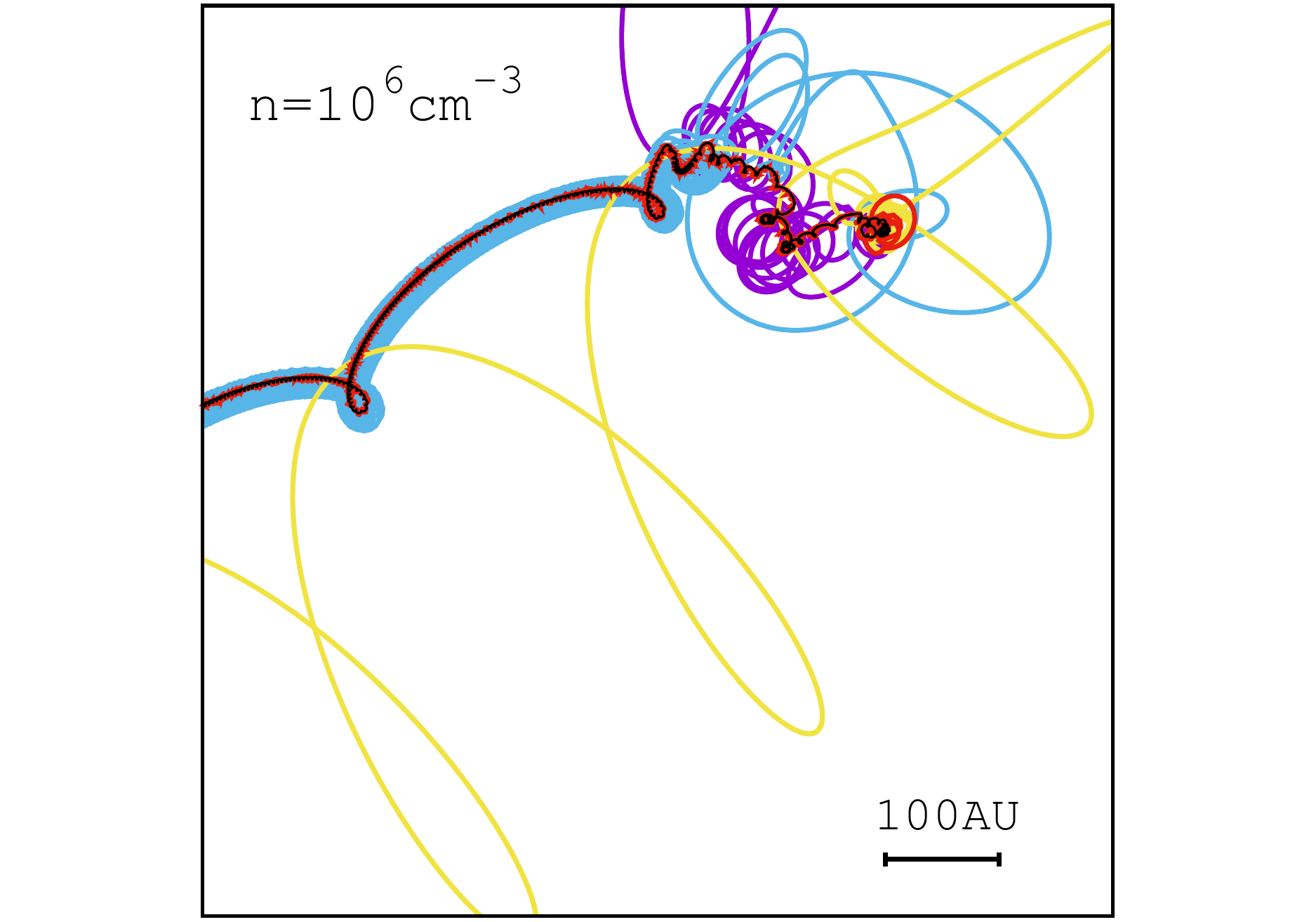}}		
		\caption{
			Trajectories up to $t = 2000~{\rm yr}$ for 5-body simulations with identical
			initial conditions but different densities: $n_{4}$ (left panel) and $n_{6}$ (right panel).
			Despite having the same starting point, the trajectories
			of the 5 stars evolve quite distinctly in the two backgrounds due to the different magnitudes of
			dynamical friction. 
			The two simulations for different bound systems: a binary
			for $n_{4}$ (black+yellow line) and a quartet for $n_{6}$
			(black+red+blue+yellow lines). }
		\label{fig:4}
	\end{figure*}
	
	\begin{figure*}
		\centering
		{\includegraphics[width=7.5cm]{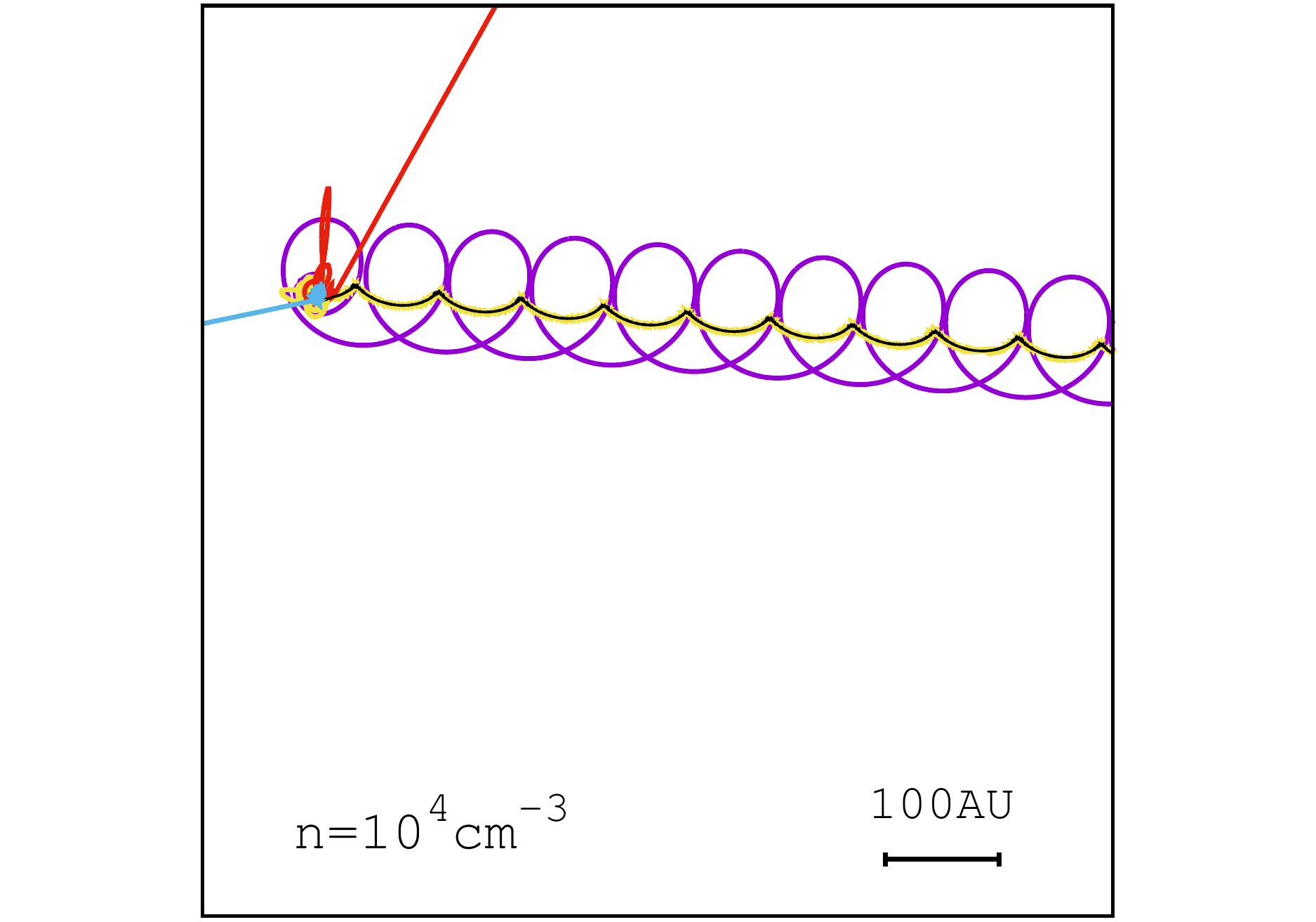}}
		{\includegraphics[width=7.5cm]{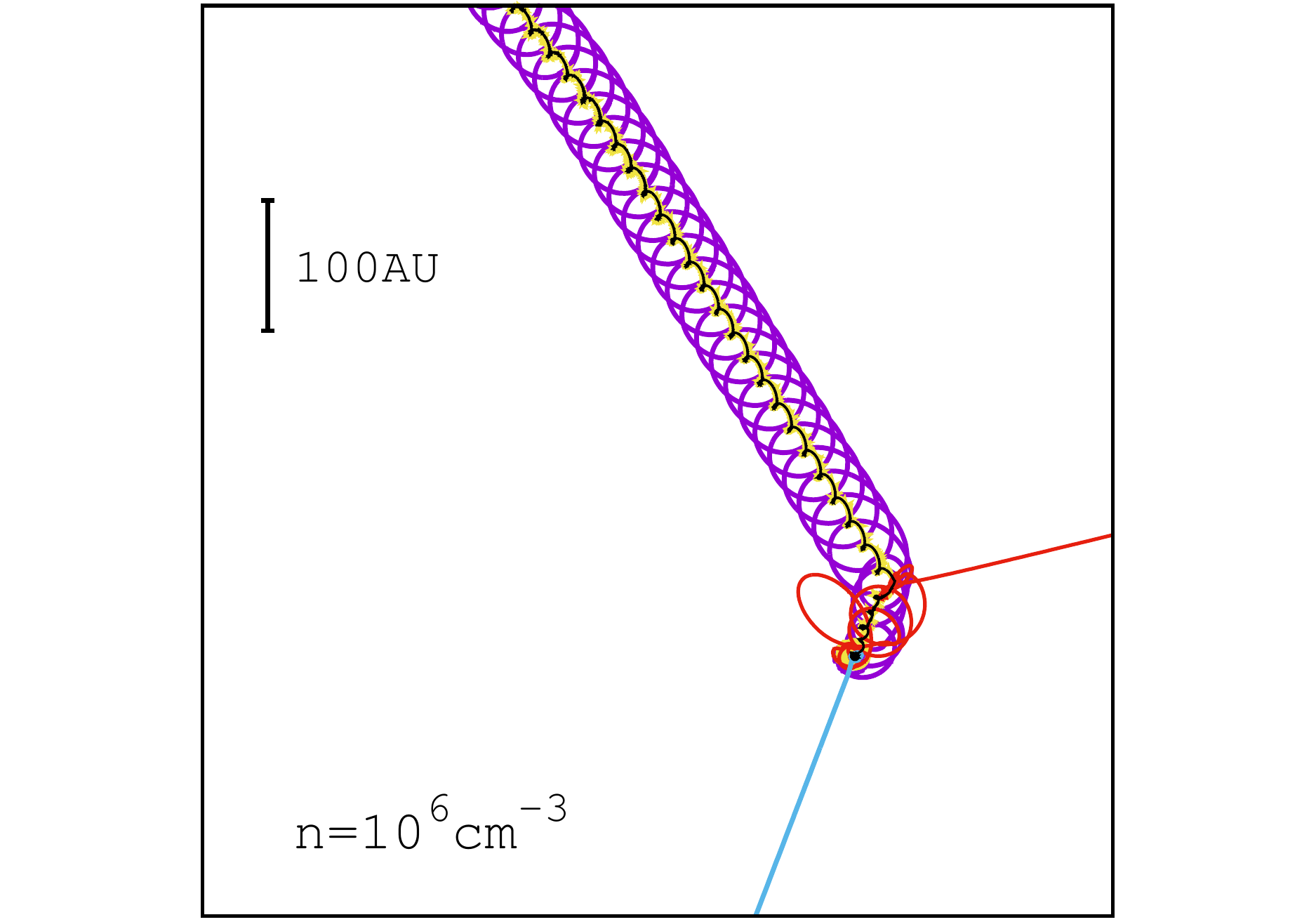}}		
		\caption{
			Trajectories of two 5-body simulations with identical initial conditions
			but different ambient gas densities.
			The simulations have formed triples with the identical stars (black+yellow+purple lines; contrast with \ref{fig:4})
			at $t = 1000\yr$, despite differences in the trajectories of each star and the center of mass.
		}
		\label{fig:3}
	\end{figure*}

	Figure~\ref{fig:4} shows two sample trajectories 
	of 5 stars with low number density (left panel) and high number density
	(right panel) after 1000 yr. They were given identical initial conditions
	for the run, but their trajectories have developed
	differently. In Figure~\ref{fig:3}, at 800 yr, even though the same stars
	form a triple and the same star is ejected for both number densities, 
	their trajectories are clearly different.

	There is no noticeable difference in the overall results 
	between the $n_4$ and $n_6$ runs---quantities such as 
	$\langle a_{t=5000~{\rm yr}}\rangle$, the dynamical friction time scale
	$\tau_{\rm df}$, the total mass of the most compact binary (or
	which star forms the compact binary with ${\rm S}_{1}$) as summarized
	in Table \ref{tab:tab2}. For both number densities,  the
	companion star of the binary is typically the third most massive star
	${\rm S}_{3}$.
	
	To sum up, we find that $P_{\rm HMXBc}\sim 7\%$ of our simulations form HMXBc, regardless
	of the gas density value.
	Normalized to the total stellar mass in the simulations, the number
	of HMXBc formed per stellar mass is
	$F_{\rm HMXB}\approx 4.6\times 10^{-4}\Msol^{-1}$.

	\subsection{10-body simulations}
	
	\label{subsec:10-body}

	We now explore several different configurations for the star group,
	and run several sets of simulations with 10 stars (instead of 5) with 
	$n=10^{6}~{\rm cm}^{-3}$. These are: (1) 10-body version of the
	calculation presented above;
	(2) head-on crash of two star groups containing 5 stars each;
	(3) a close encounter and subsequent inspiral and merger of two star groups
	containing 5 stars each. 
	
	We generate stellar masses in the same way as for the 5-body case,
	but with a larger value for the parameter $M_{\rm max}=300\Msol$.
	We have run each simulation for 500 yr. The
	results of these simulations are summarized in Table \ref{tab:tab3}.
	We briefly discuss each one, as follows.

			\begin{table}
				\centering
				\setlength\extrarowheight{2pt}
				\begin{tabulary}{0.6\textwidth}{c c c c c c}
					\hline
					Scenario & 1 & 2 & 3 & 4A & 4B \\
					\hline
					runs 													      & 54	   	  &	 30	  	& 30	 	 & 	30	  	&	30	\\
					$\langle a_{t=500\yr}\rangle$   $[\AU]$     & 1.0 	  	&  1.1	  & 0.90	  &	1.8	  	 & 	 1.6 \\
					$\langle a_{\rm{B}_{12}}\rangle$ $[\AU]$  & 1.6	  	  &  1.5	 &	1.8	  	&   2.4	 	&    2.0	  \\
					$\langle a_{\rm{rest}}\rangle$  $[\AU]$ 	  &  0.72   &  	0.15  & 0.42	 & 1.2		&	0.23  \\
					$P_{\rm{HMXBc}}$  									&  0.33	   &  0.33	&  0.27		&	0.13  & 	0.27  \\
					$F_{\rm{HMXBc}}$ [$10^{-4}\Msol^{-1}$] &    15	   &  11	  &  9.0	  &	4.2      & 	8.4  \\
					\hline
				\end{tabulary}
				\caption{Summary of the results from 10-body simulations. We have
					considered five
					different scenarios.
					Scenario 1 simulated the 10-body version of the 5-body calculation,
					i.e. an isolated star group.
					The other scenarios all involve collisions of two 5-body star groups.
					Scenario 2 is a head-on collision of two coplanar, co-rotating star groups.
					Scenario 3 also collided two groups of stars with co-rotating orbital planes,
					but with an impact parameter comparable to
					the sizes of the groups, which results in an inspiral and eventual merger.
					Scenarios 4A and 4B are similar to scenario 3, except that
					the orbital planes of the colliding star groups had
					mutual inclinations of $45^\circ$ and $135^\circ$, respectively.
				}
				\label{tab:tab3}
			\end{table}

	\subsubsection{Scenario 1: 10-body group in isolation}
	\label{subsubsec:Scenario1}
	We set up the simulations as in the 5-body calculations, but with 10 stars.

	We find that $\langle a_{t=500{\rm yr}}\rangle=1.0~{\rm
		AU}$. Interestingly, there are 18 out of 54 cases in which
	$a<a_{{\rm RL}}$. There is a large difference in scale between
	$a_{{\rm B}_{12}}$ and $a_{{\rm rest}}$, where $a_{{\rm B}_{12}}$ is
	the semi-major axis of ${\rm B}_{12}$ (binary made up of the two most
	massive stars) and $a_{{\rm rest}}$ is the semi-major axis of the
	binary stars other than ${\rm S}_{1}$ and ${\rm S}_{2}$. 
	This is a common feature of the 10-body simulations:
	they often end up with triples whose inner
	binary is $B_{\rm rest}$ while the outer binary is $B_{12}$. Our
	simulations yield $\langle a_{{\rm rest}}\rangle =0.72~{\rm AU}$ and
	$\langle a_{{\rm B}_{12}}\rangle=1.6~{\rm AU}$. Also note that, in 14
	out of 18 HMXBc, the binary is ${\rm B}_{{\rm rest}}$ (i.e. it is not
	made up of the two most massive stars).
	
	We find that a HMXBc forms in a larger fraction of these simulations
	than in the 5-body case, $P_{\rm HMXBc}=0.33$,
	for the obvious reason that there are more stars. 
	Per unit stellar mass in the simulations, the number of HMXB candidates
	is $F_{\rm HMXB}=1.5\times 10^{-3}\Msol^{-1}$,
	which is a factor $\approx 3$ higher than we found for the 5-body case.

	\subsubsection{Scenario 2: Collision between two 5-star groups -- head-on collision}
	\label{subsubsec:Scenario2}
	
	In scenario 2, two groups of 5 stars collide head-on. The initial conditions for two groups of 5 stars are generated randomly in the same way as for 5-body simulations. Their separations are two to three
	times their sizes and the mutual inclination is zero. 
	The initial relative speed of the groups is roughly
	the speed of sound and the center of mass of one group is set to move
	directly toward the center of mass of the other group.
	
	We find that prior to colliding, each group forms a compact binary of
	type ${\rm B}_{12}$ (the most and second-most massive
	star). When the two groups collide, those two binaries
	that existed before the collision were broken and the two most
	massive stars of each group form a new compact binary with high
	chances. The
	average $\langle a_{t=500~{\rm yr}}\rangle$ is $1.1~{\rm AU}$, but
	$\langle a_{{\rm B}_{12}}\rangle=1.5~{\rm AU}$ and $\langle
	a_{{\rm rest}}\rangle =0.15~{\rm AU}$, meaning that the most compact
	binaries are not formed from the most massive stars.
	The shorter average separations may be a result of a larger
	number of early 3-body scatterings, which act to harden 
	the group as a whole.
	
\begin{figure*}
			\centering
			{\includegraphics[width=4.8cm]{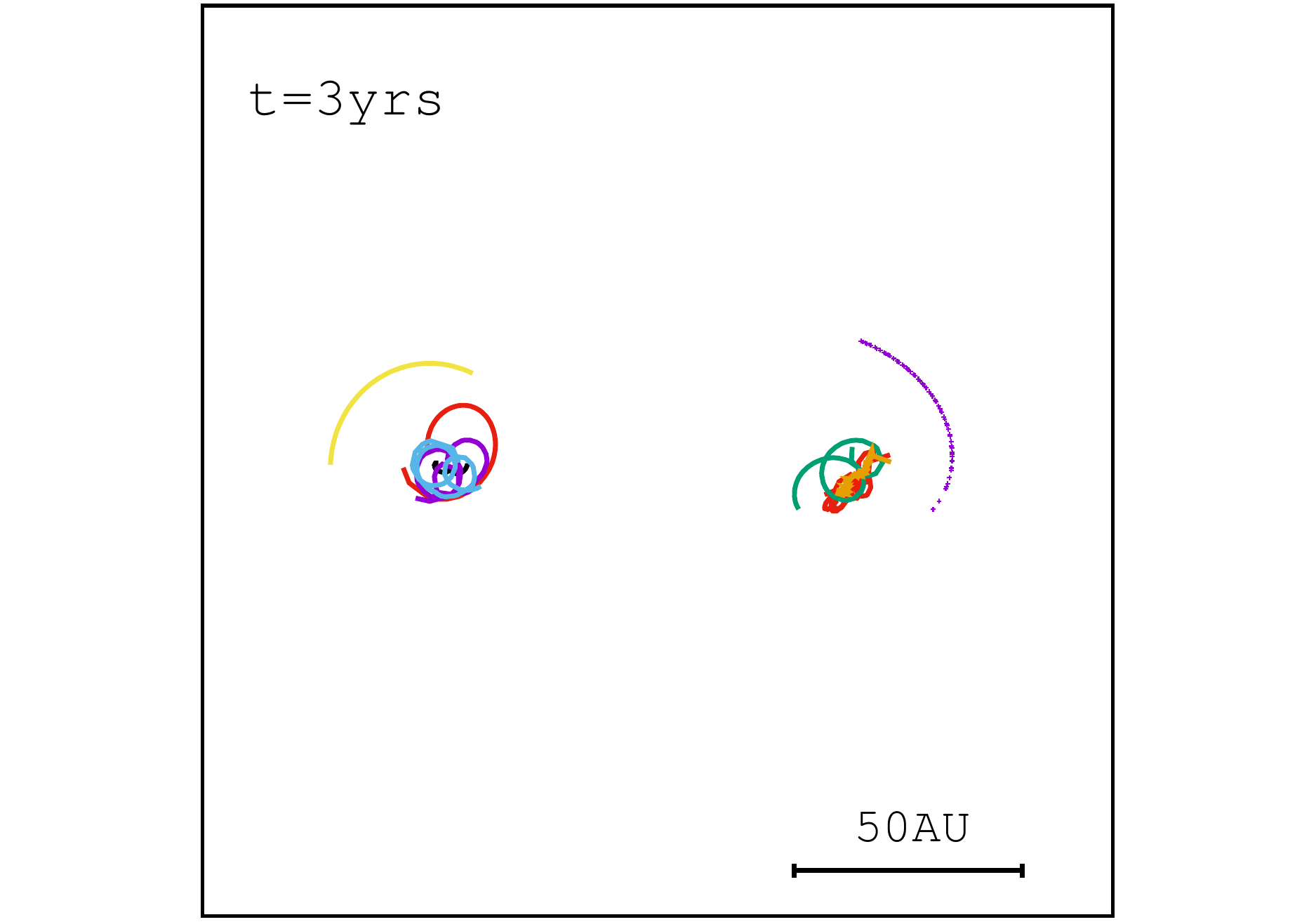}}
			{\includegraphics[width=4.8cm]{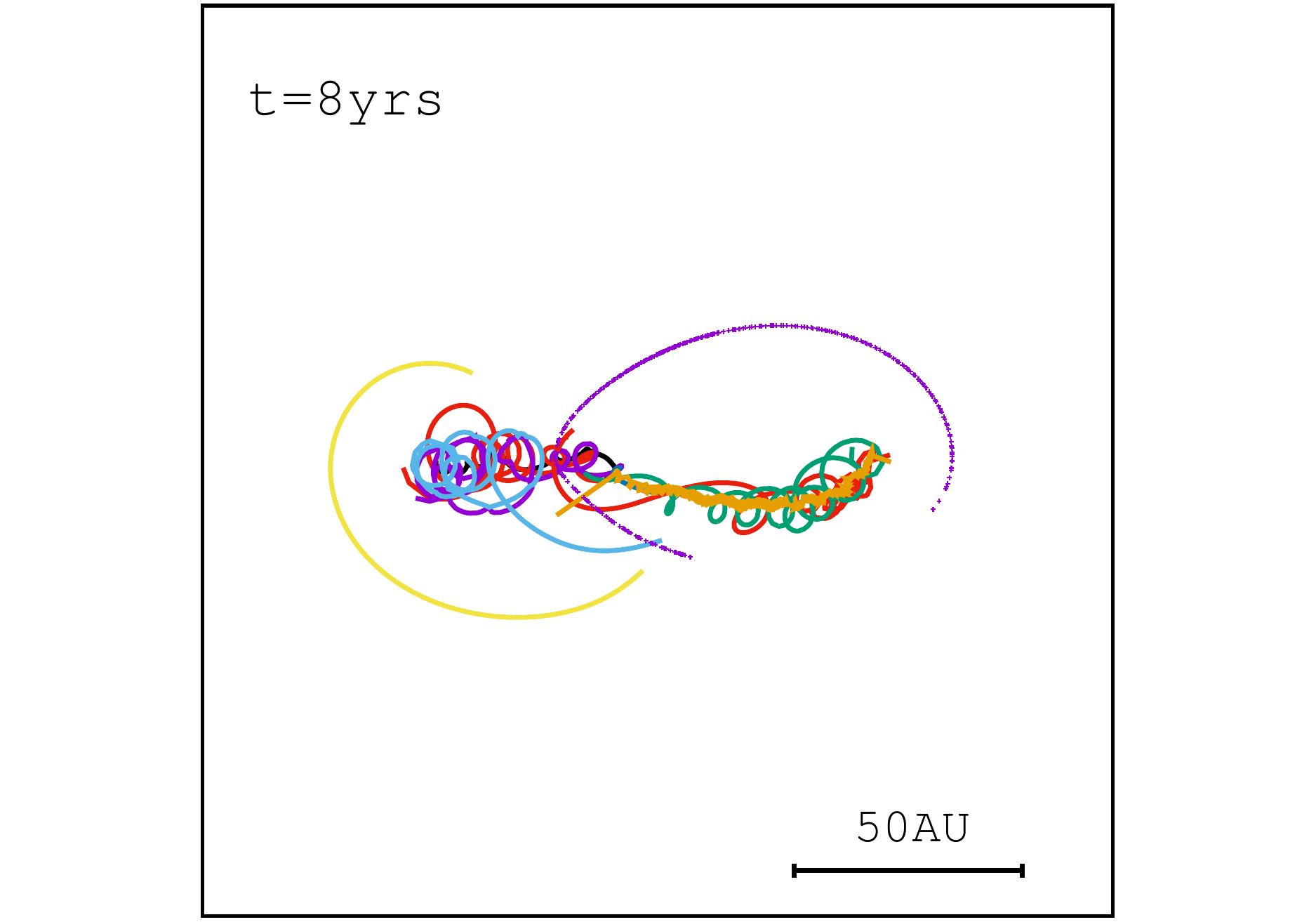}}
			{\includegraphics[width=4.8cm]{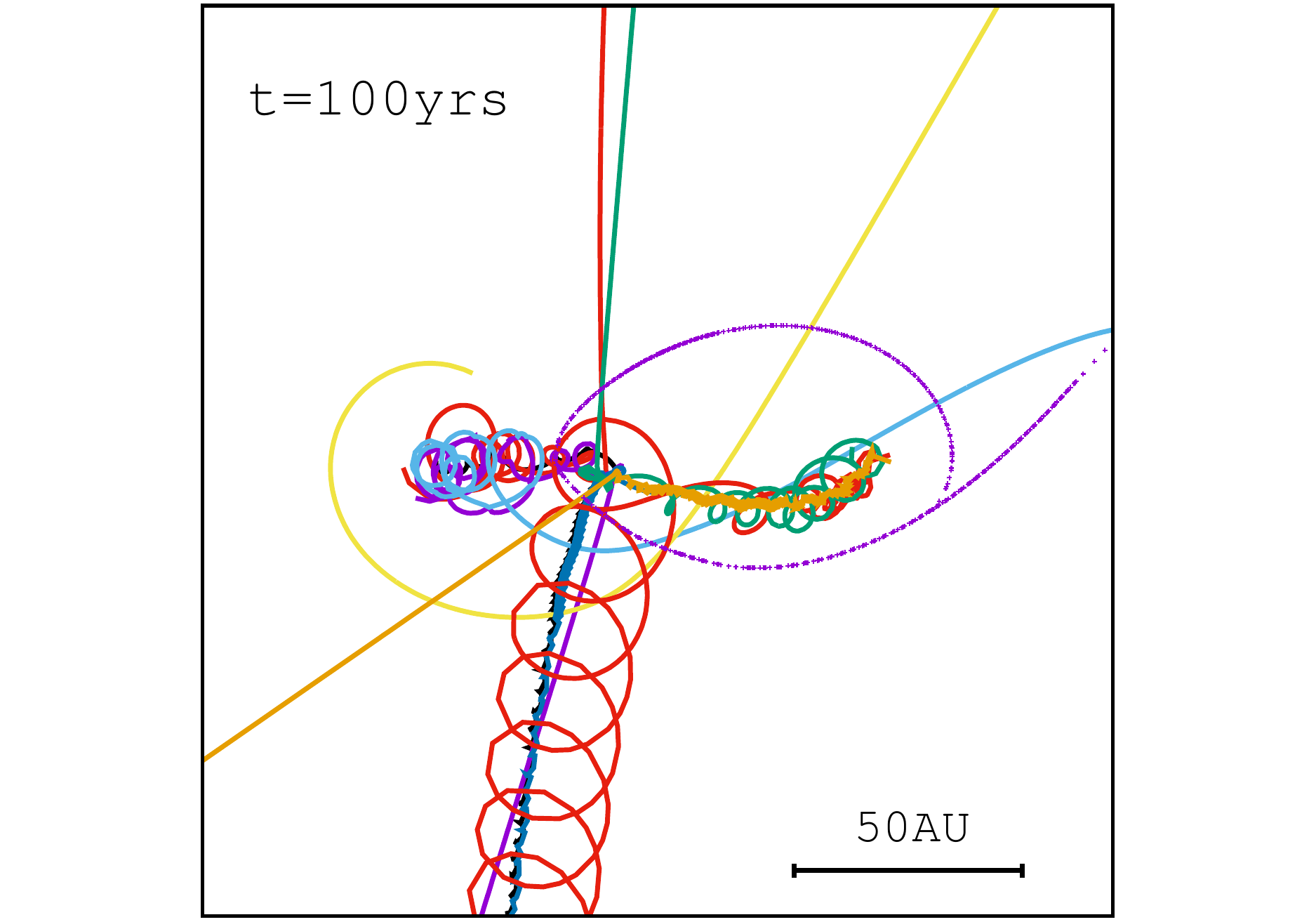}}				
			\caption{ 10-body head-collision (scenario 2). In
				these sample trajectories, five stars in each group are in nearly
				Keplerian motion. The groups move towards each other
				with a relative velocity of $\sim c_{\rm s}$ (left
				panel). At $t\approx 8$~yr (middle panel), the two haloes
				begin to merge. During the merger, all ten stars undergo
				stellar scatterings (right panel). One triple (black+dark blue+red lines)
				has been formed and it is moving in -y direction. In these
				head-on collisions, it is likely for multiple systems that
				existed before merging to be broken, while a few new
				multiple systems are formed.}
			\label{fig:10}
		\end{figure*}

	HMXBc form in 10 out of 30 runs, and they are not of type ${\rm B}_{12}$.
	However, we find a rate of HMXB formation per stellar mass
	$F_{\rm HMXB}=1.1\times 10^{-3}\Msol^{-1}$;
	this is higher than in the 5-body case and comparable to Scenario 1 above.
	Sample trajectories 
	for one of the simulations of scenario 2 are depicted in Figure~\ref{fig:10}.

		\begin{figure*}
			\centering
			{\includegraphics[width=4.8cm]{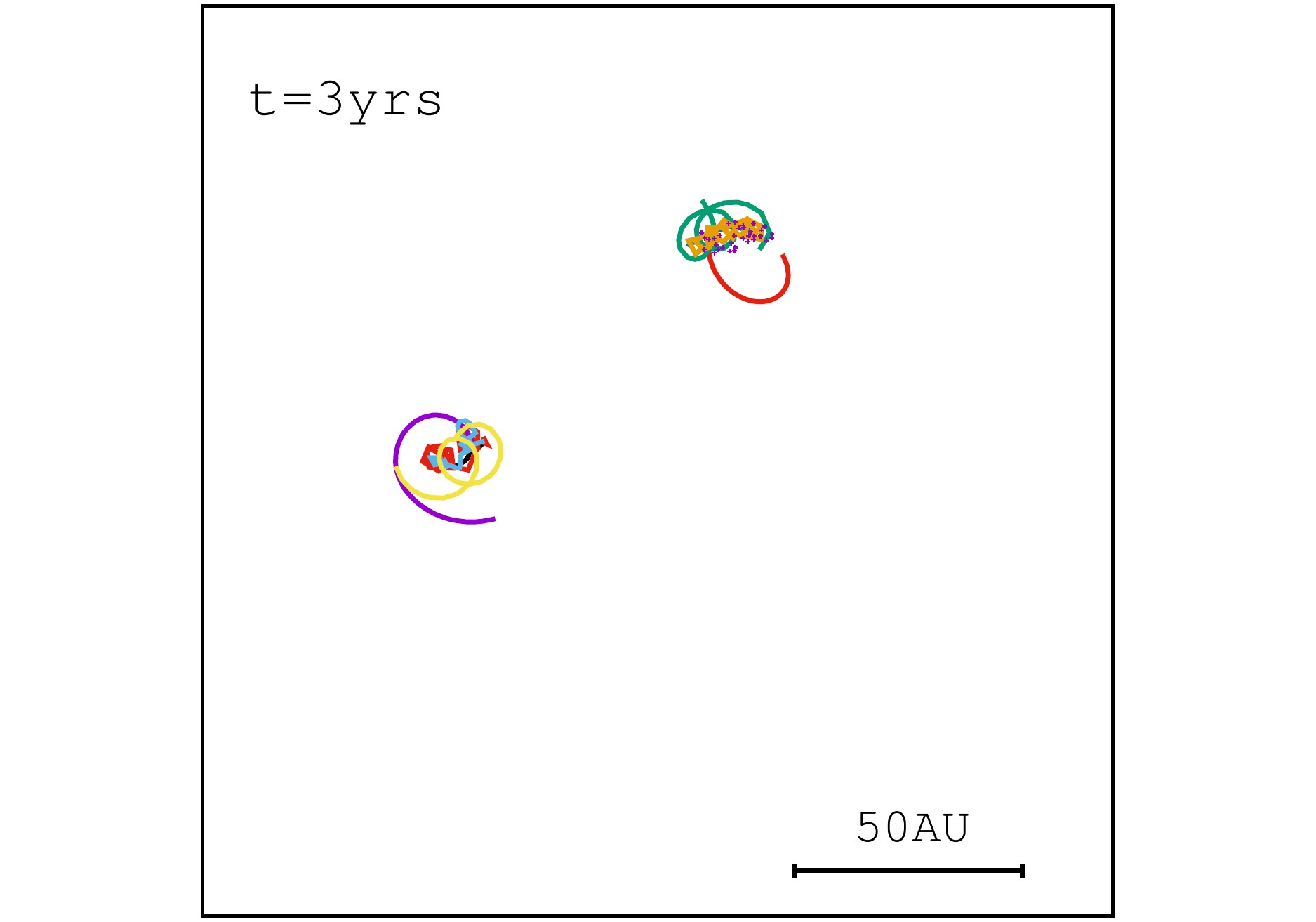}}
			{\includegraphics[width=4.8cm]{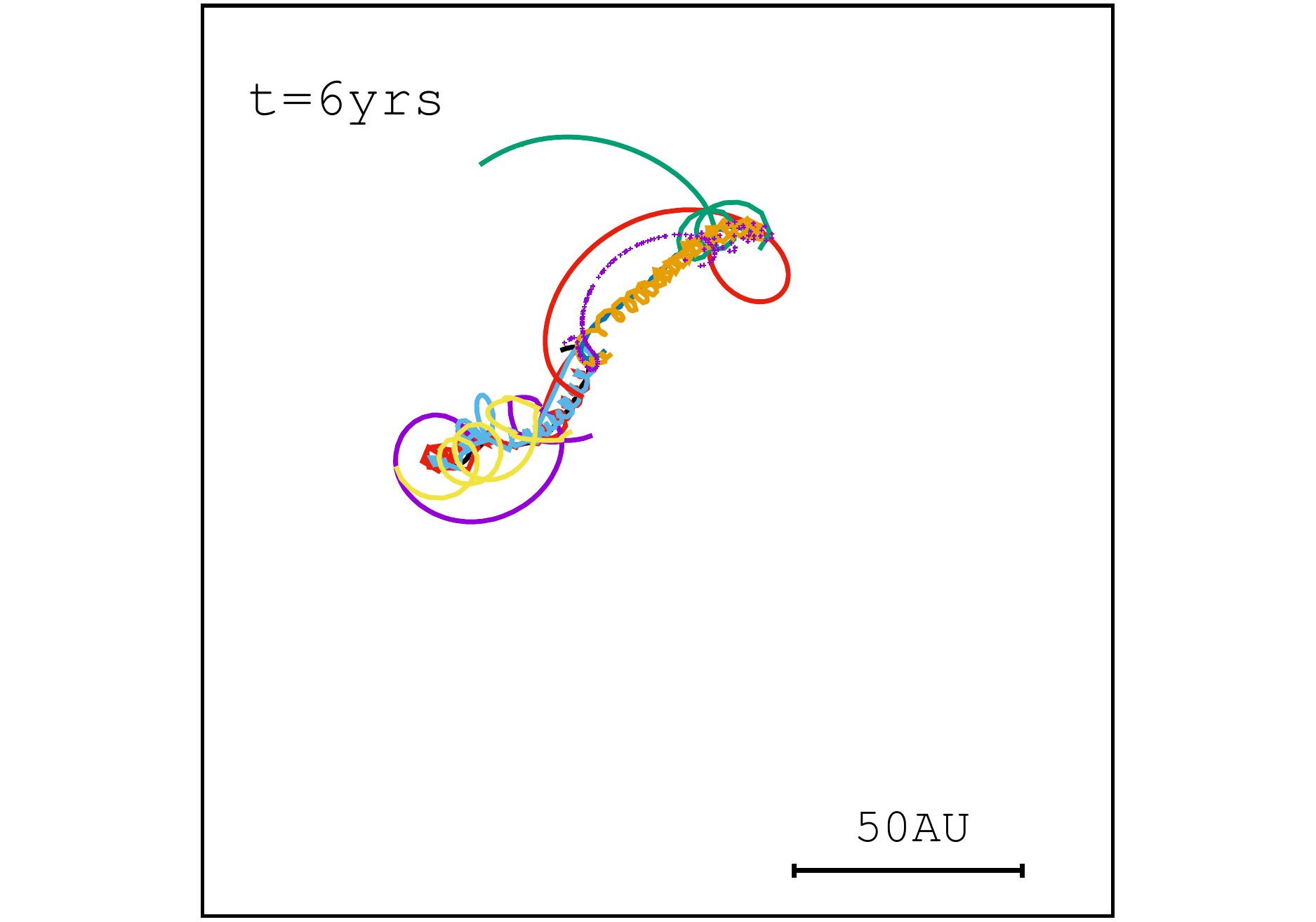}}
			{\includegraphics[width=4.8cm]{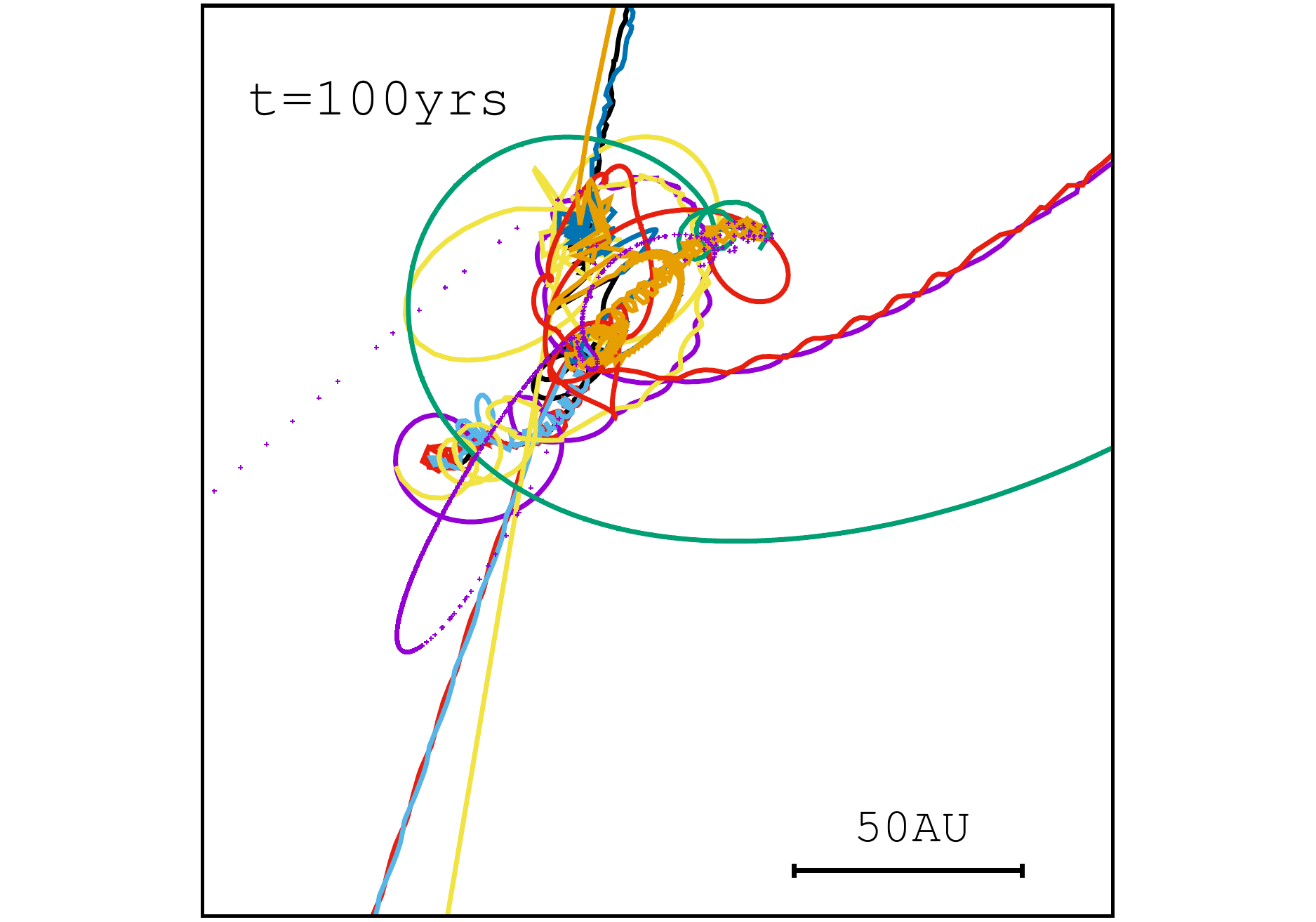}}				
			\caption{10-body spirally merging case
				(scenario~3). As with scenario 2, two groups of 5 stars
				are set on a collision cours. However, in this scenario the impact parameter is
				$\sim 2-3$ times the size of groups, whereas in
				scenario~2 it is set to be zero. The groups are
				approaching each other at the relative velocity of $c_{\rm s}$
				(left panel). At $t=8$~yr, the groups are about to
				merge (middle panel). After some time (right panel),
				two binaries are ejected (red and purple lines
				in the $x$-direction and black and blue lines in the $y$-direction). In this
				sample case, the most compact binary is the one ejected in
				the y direction (black+blue lines) with $a=0.8$~AU at
				$t=100$~yr.  A difference between this scenario and the head-on collision
				is that compact systems are more likely to survive
				the merger.}
			\label{fig:11}
		\end{figure*}

	\subsubsection{Scenario 3: Collision between two 5-star groups -- spirally merging case}
	\label{subsubsec:Scenario3}
	
	We have used the same orbital parameters for the two groups as in
	Scenario 2, except that we now set the impact parameter to be of order
	the size of the group, whereas it was set to zero in Scenario 2. So they merge with a spiral motion.
	
	We find that  $\langle a_{t=500~{\rm yr}}\rangle = 0.90\AU$, $\langle
	a_{{\rm B}_{12}}\rangle=1.8\AU$ and $\langle a_{{\rm rest}}\rangle
	=0.42\AU$. The average separation lies between what we find in Scenarios 1 and 2.
	This can be interpreted as being due to the fact that these simulations
	(in which the two groups merge gradually via inspiral)
	have more close 3-body interactions than in Scenario 1 (in which
	10 stars in quasi-Keplerian orbits evolve in isolation)
	but fewer such interactions than in Scenario 2
	(in which the two groups merge head-on).

	HMXBs form in 8 out of 30 runs and, as with Scenarios 1 and 2,
	none of the HMXBs are made up of the two most massive stars.
	We find a similar HMXB formation rate per stellar mass,
	$F_{\rm HMXB}=9.0\times 10^{-4}\Msol^{-1}$.
	Sample trajectories from one of the
	simulations for Scenario 3 are shown in Figure~\ref{fig:11}.

	\begin{figure*}
		\centering
		{\includegraphics[width=4.8cm]{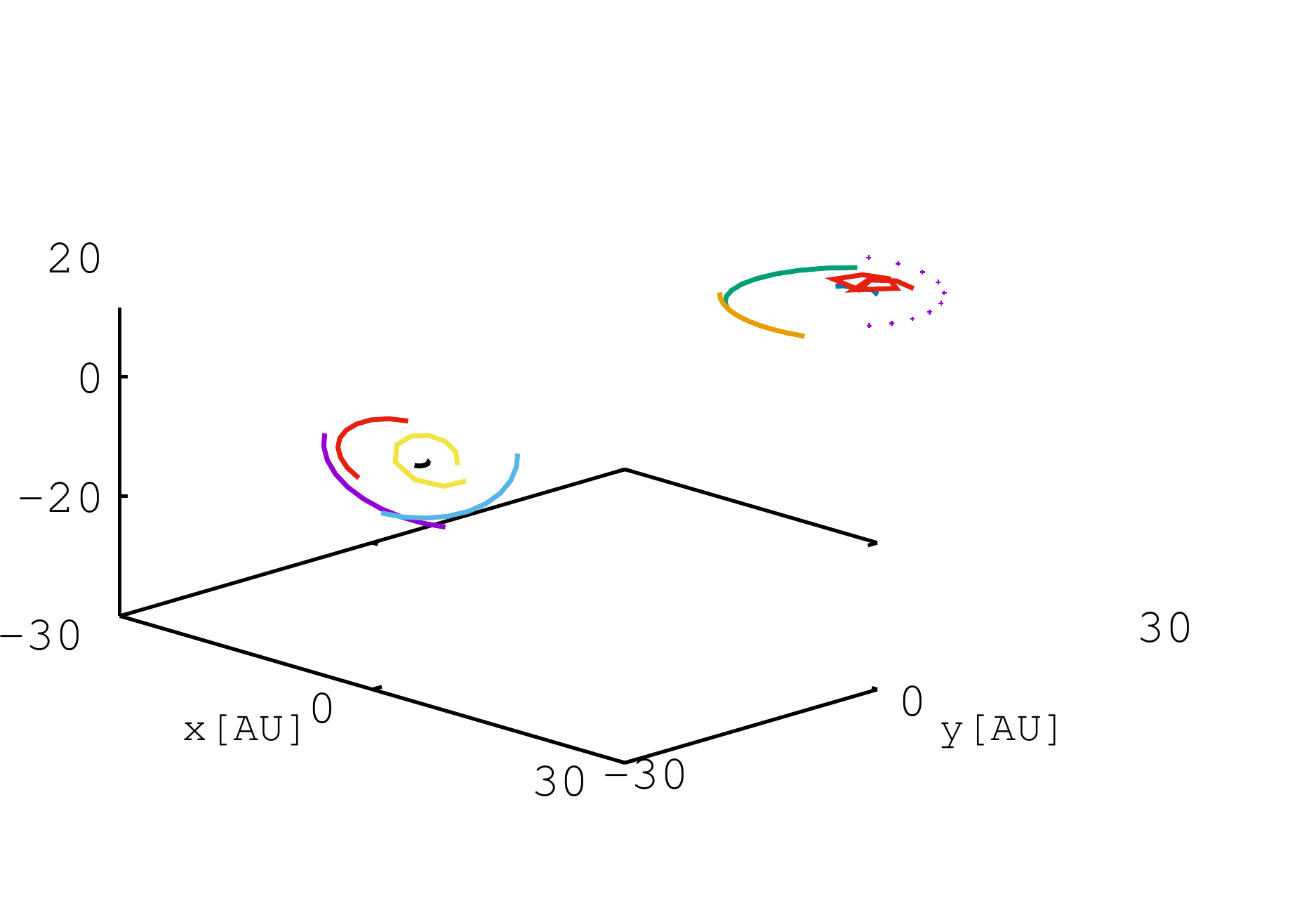}}
		{\includegraphics[width=4.8cm]{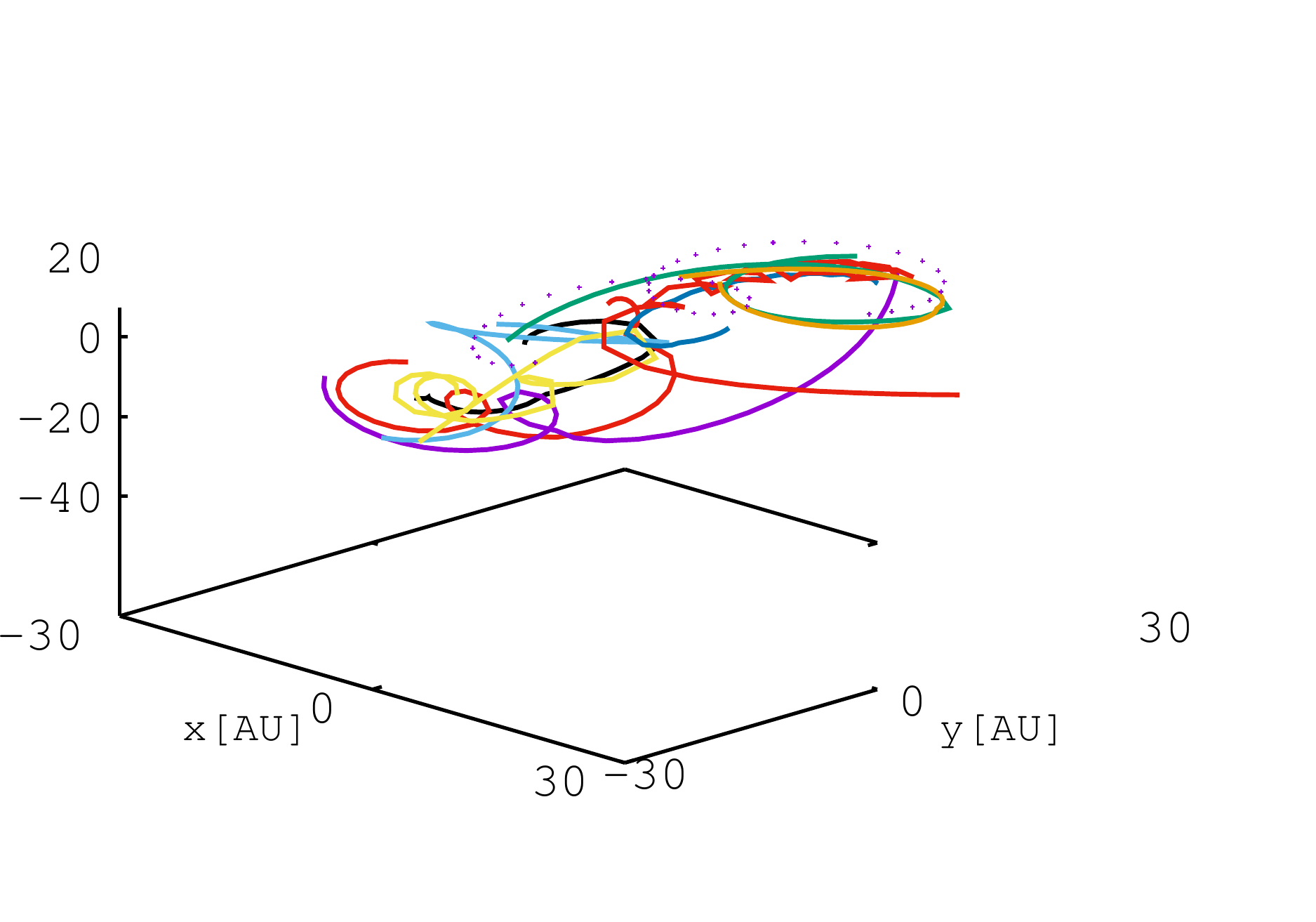}}
		{\includegraphics[width=4.8cm]{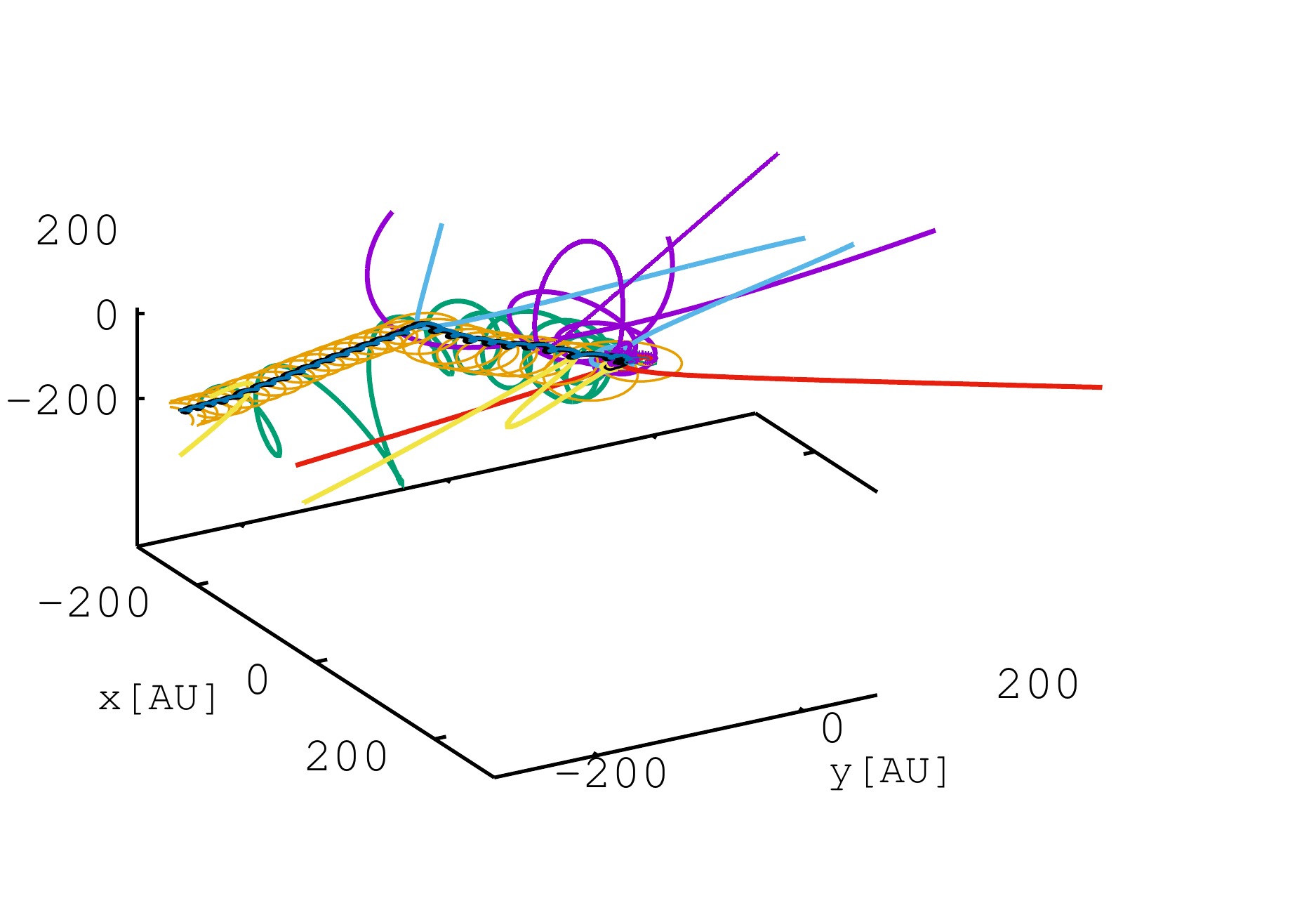}}				
		\caption{Trajectories for scenario 4A, a collision between two 5-body groups, each
			on quasi-Keplerian orbits, but with the orbital plane
			of the two groups tilted with respect to each other
			at an inclination  $i= 45^{\circ}$.
			The setup is the same as scenario 3,
			except for the mutual inclination of the orbital planes of the colliding
			star groups.
			The left panel shows the two groups on a collision course.
			At $t=4\yr$, the halos are about to
			merge (middle panel). After $t=10\yr$ (right panel),
			a triple having the most compact binary are ejected [black and blue lines (inner binary) and brown line].}
		\label{fig:12}
	\end{figure*}

	\begin{figure*}
		\centering
		{\includegraphics[width=4.8cm]{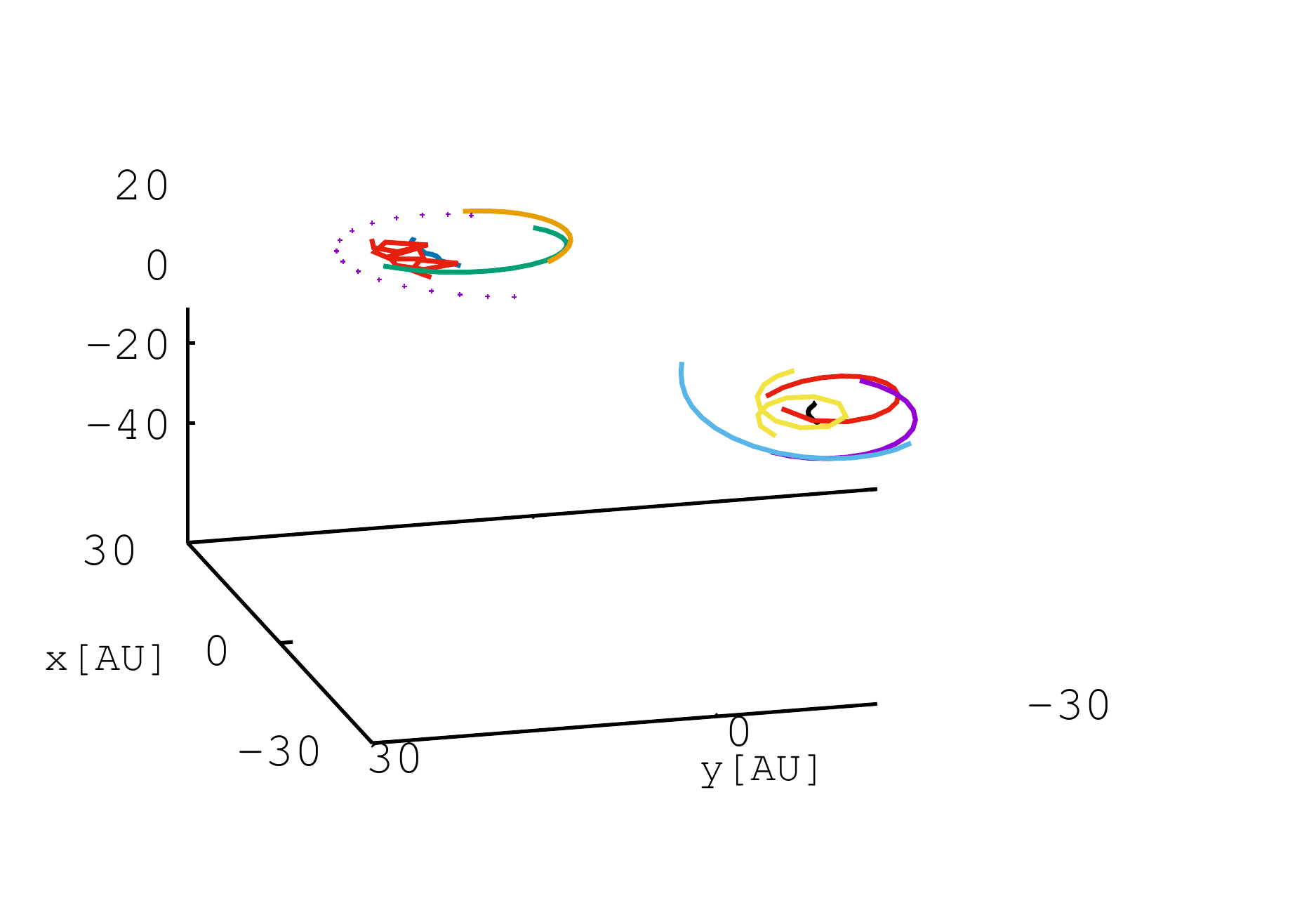}}
		{\includegraphics[width=4.8cm]{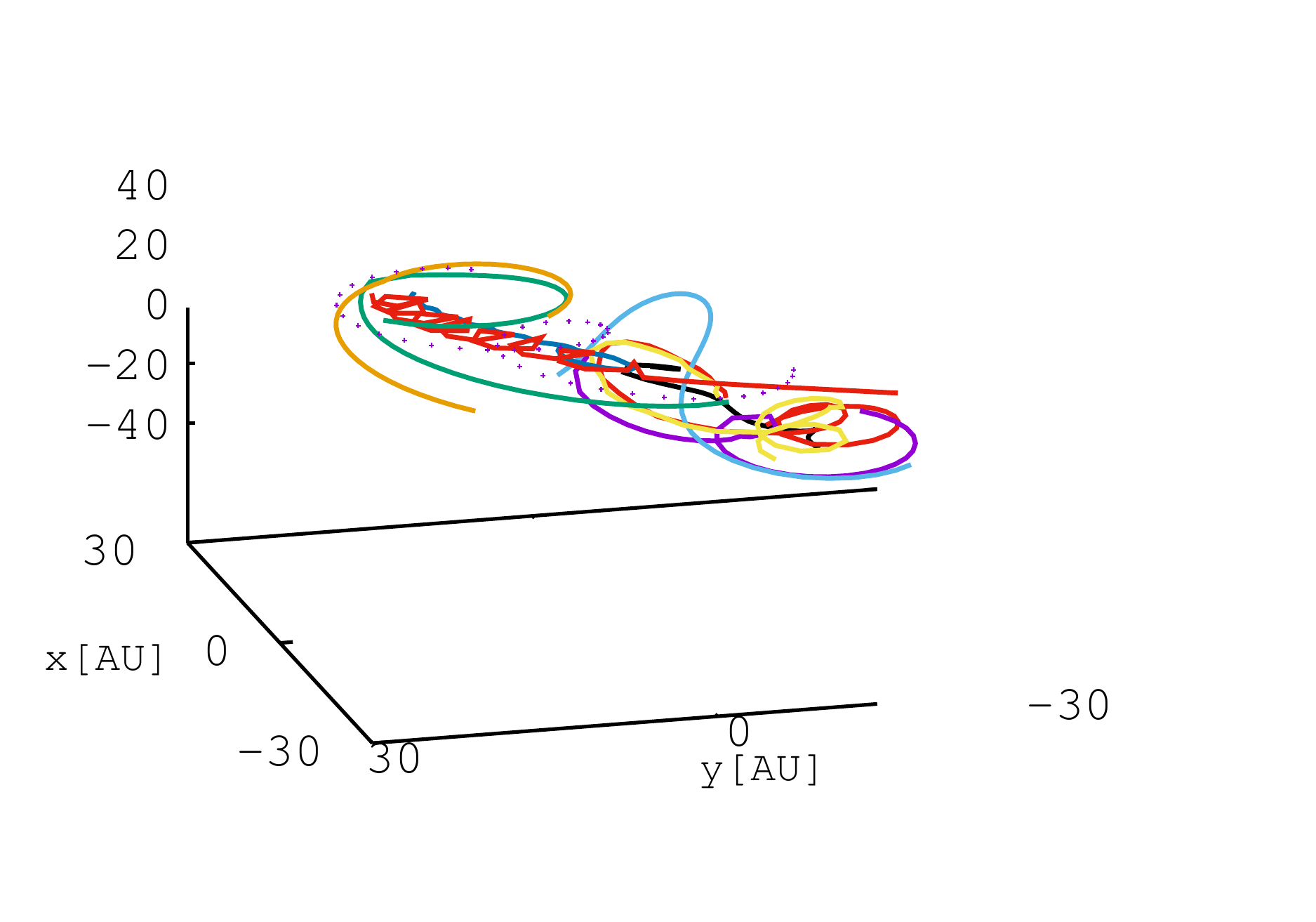}}
		{\includegraphics[width=4.8cm]{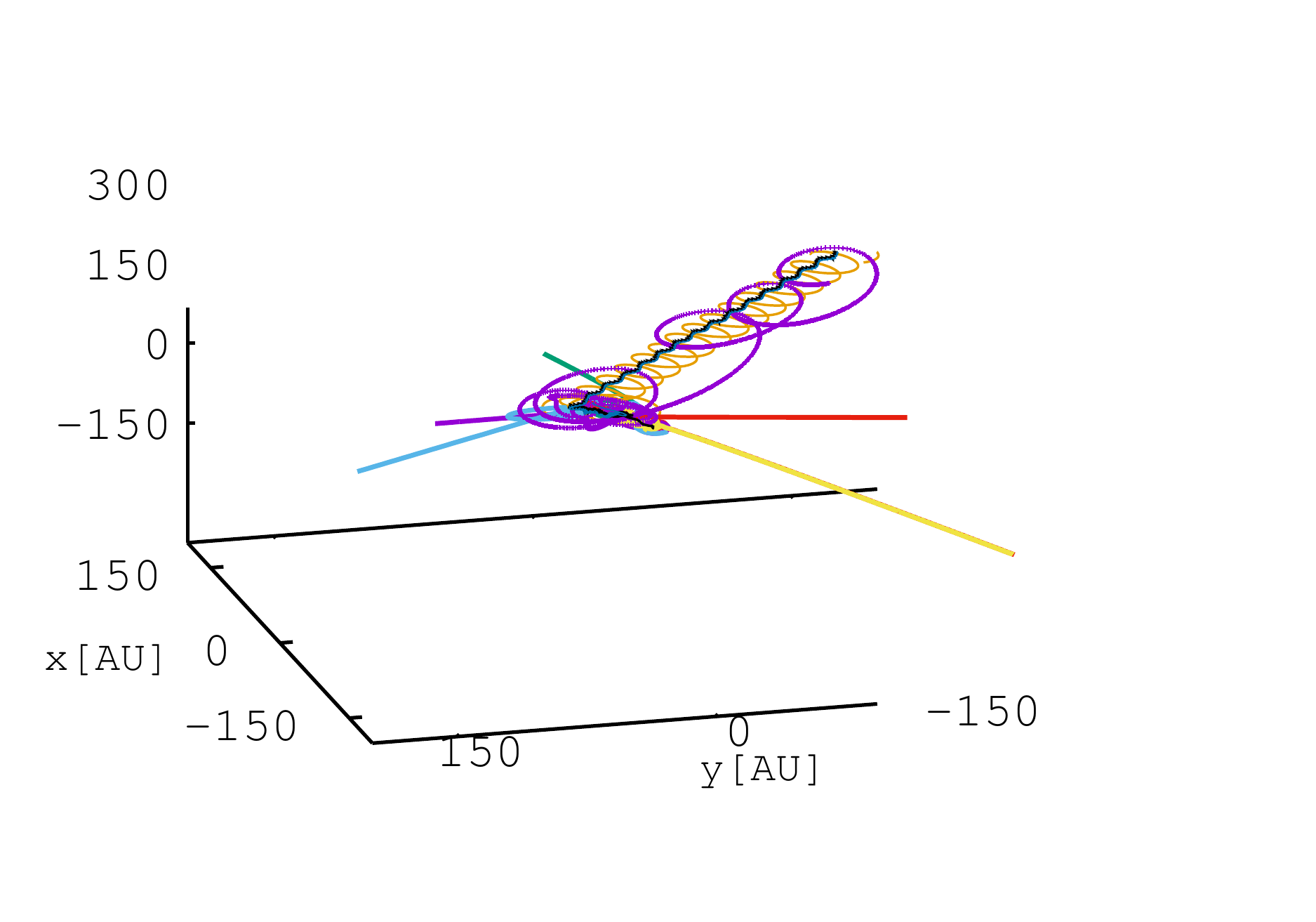}}				
		\caption{Trajectories for scenario 4B---the same as scenario 4A (Figure~\ref{fig:12}),
			but with the two groups nearly counter-rotating with respect to each other
			($i=135^{\circ}$). At $t=4\yr$,
			the groups are about to merge (middle panel) and after $t=10\yr$ (right panel),
			a quartet containing the most compact binary is ejected toward the upper right
			of the panel [black+blue lines (inner binary), brown and thick purple lines].}
		\label{fig:13}
	\end{figure*}
	\subsubsection{Scenarios 4A and 4B: Collision between two 5-body groups -- spirally merging case with inclinations of 45 degrees and 135 degrees}
	\label{subsubsec:Scenario4}
	In these two scenarios, the difference compared with those previous scenarios is that the orbital plane of one star group is tilted,
	or non-zero mutual inclination $i$.
	We take the inclination at $i=45^\circ$ (nearly co-rotating) for scenario 4A,
	and $i=135^\circ$ (nearly counter-rotating) for scenario 4B.
	The groups are initially placed at a separation of two to three
	times their sizes, and set in motion at the same speeds
	as for the inspiral case (scenario 3).
	
	We find that $\langle a_{t=500\yr}\rangle = 1.8\AU$,
	$\langle a_{\rm{B}_{12}}\rangle=2.4\AU$,
	and $\langle a_{\rm{rest}}\rangle=1.2\AU$ for scenario 4A.
	In scenario 4B, we find
	$\langle a_{t=500\yr}\rangle =1.6\AU$,
	$\langle a_{\rm{B}_{12}}\rangle=2.0\AU$,
	and $\langle a_{\rm{rest}}\rangle=0.23\AU$.
	In $70\%$ of the runs for both scenarios, the two most massive
	stars form the most compact binary.
	Stellar binaries end up with somewhat closer separations in the nearly counter-rotating case,
	due to the fact that the net angular momentum of the merged star group is smaller.
	Indeed, we find a total of four HMXBc across all the $i=45^\circ$ simulations,
	and eight in a same number of $i=135^\circ$ simulations.
	
	For the same reason, we find a larger fraction of
	HMXB candidates per stellar mass simulated in the nearly counter-rotating case
	($F_{\rm HMXB}=8.4\times 10^{-4}\Msol$)
	compared to the nearly co-rotating case ($F_{\rm HMXB}=4.2\times 10^{-4}\Msol$).
	The overall formation rate of HMXB candidates is lower for both cases than the
	cases in which all the stellar orbits were nearly coplanar (Scenarios 1, 2 and 3),
	plausibly due to the additional degree of freedom in the stellar orbits.
	Still, the value of $F_{\rm HMXB}$ is within a factor of a few for all of our simulations.
	
	Sample trajectories from runs for scenarios 4A and 4B are depicted in
	Figure~\ref{fig:12} and Figure~\ref{fig:13}, respectively.

	\section{Discussion}
	\label{sec:discussion}
	
	\subsection{Binary evolution and formation of HMXB candidates}
	\label{subsec:Binaryevolution}

	Our simulations show that if protostellar clouds fragment and form stars in close
	groups on scales of $\simgt 10\AU$, as in \cite{greif11}, then a small fraction
	of groups can form HMXBs. We briefly discuss the dynamics of
	HMXB formation in our simulations, then move on to discuss the
	astrophysical implications of our findings.

	According to our simulations,
	scatterings play a major role in making a compact binary.
	The background potential and the dynamical friction play a secondary role,
	by allowing the most compact binary (or triple) to remain near the center
	of mass of the halo and for other stars to return and scatter again and again.
	
	We find that on average,
	the number of HMXB candidates formed per stellar mass, $F_{\rm HMXB}$,
	depends on the number and orientation of close 3-body encounters.
	Our results indicate that if fewer number of stars are ejected during scattering and their interactions are coplanar, $F_{\rm HMXB}$ may be somewhat higher.
	It is important to note that the value of $F_{\rm HMXB}$ does not vary by more than a factor $\approx 3$ throughout all scenarios. We interpret this lack of a significant variation in $F_{\rm HMXB}$,
	for such a diverse set of initial conditions and ambient gas densities,
	to mean that our values are not far from the one that results
	from similar stellar encounters in nature.

	\subsection{The effect of migration on the formation of HMXBs}\label{subsec:migration}
	
	The migration process could be one another way to harden the stellar group through a gaseous disk.
	The migration could occur as the protostars form---\cite{Greif12} 
	found significant accretion from the protostellar disk onto the most massive
	protostar, and did not follow the evolution of the system beyond this stage.
	
	We evaluate the possible role of disk migration on the separation of Pop~III
	stars by considering a steady, geometrically thin disk with an $\alpha$ viscosity
	(\cite{SSdisk}, \cite{Frank2002accretion}).
	We adopt a disk with $\alpha=0.01$ and an accretion rate
	$\dot{m}\sim 10^{-3}~\Msol\yr^{-1}$ (\cite{Tan2004a}, \cite{Tan2004b}),
	who considered the structure of accretion disks around Pop~III
	stars at high redshifts.
	
	We estimate the migration timescale $\tau_{\rm mig}$ (\cite{Syer1995}) for a representative binary system with primary mass $M_{1}=120\Msol$
	and secondary mass $M_{2}=11\Msol$, based on the mean values
	found across our simulations.

	\begin{figure}
		\centering
		{\includegraphics[width=8.9cm]{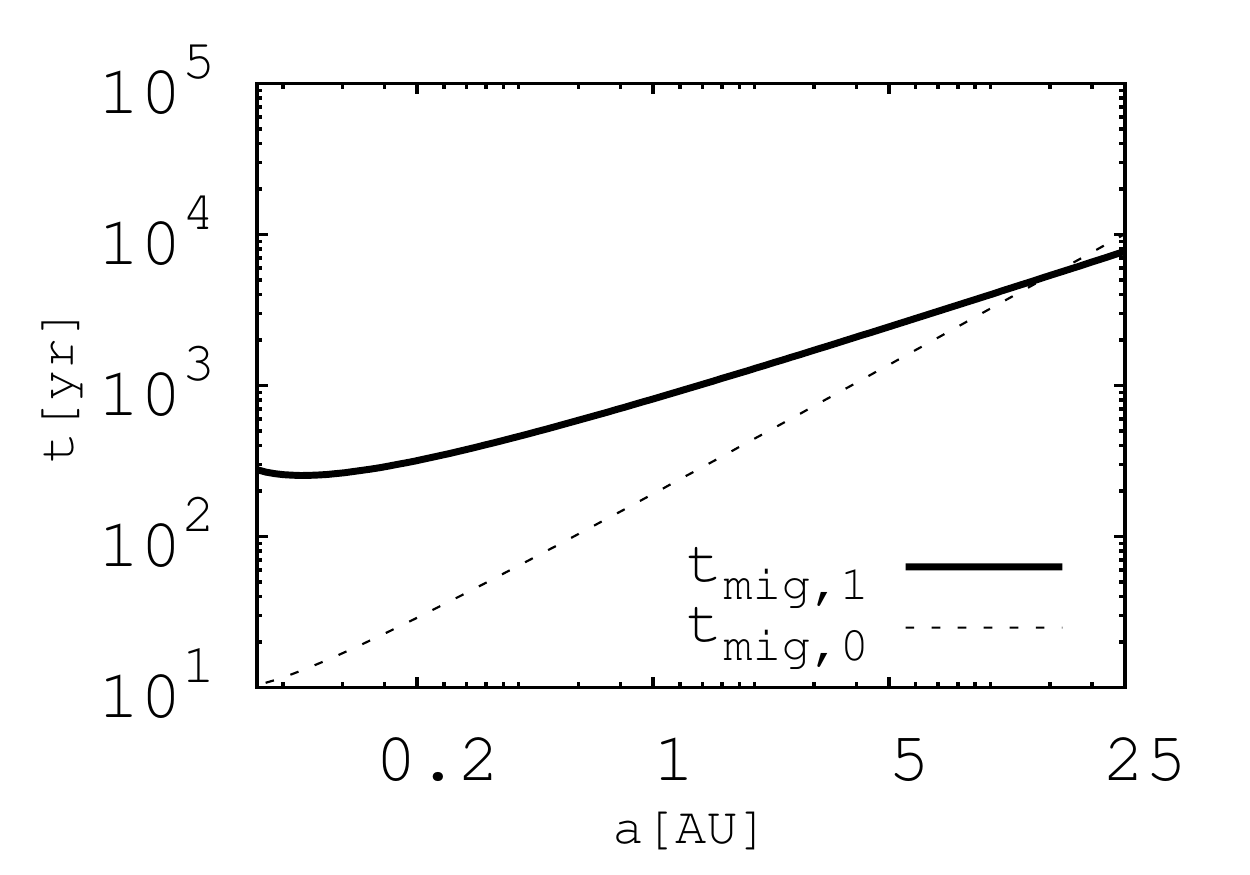}}
		\caption{
			The migration timescales for a circumbinary
			protostellar disk, based on the typical binary properties of our
			simulations. 
			The unperturbed Type II migration timescale, $\tau_{{\rm mig},0}$ (dashed line)
			shows the timescale assuming that the secondary mass is much
			smaller than the local disk mass.
			A longer timescale $\tau_{{\rm mig},1}$ (solid line) is expected if
			the secondary mass is large compared to the disk mass.}
		\label{fig:9}
	\end{figure}
	
		We plot the two migration timescales $\tau_{{\rm mig},0}$ and $\tau_{{\rm mig},1}$
		in Figure~\ref{fig:9}. The unperturbed Type II migration timescale, $\tau_{{\rm mig},0}$ (dashed line)
		shows the timescale assuming that the secondary mass is much
		smaller than the local disk mass.
		A longer timescale $\tau_{{\rm mig},1}$ (solid line) is expected if
		the secondary mass is large compared to the disk mass.
		Both of these timescales are shorter than both the typical lifetimes
	of protostellar disks, as well as the lifes of the stars
	themselves{\footnote{Note that, if the adopted value of $\alpha$
			were higher, e.g. $\sim 0.1-1$ as suggested by recent simulations
			of Pop~III protostellar disks \cite{Clark2011}, the migration
			timescale would become even shorter.}. This suggests, along with
		findings that the disks may be able to survive even
		under radiation feedback (\cite{Stacy+12},
		\cite{Hosokawa+2015}), that disk migration could 
		lead to initial stellar separations smaller than what we have assumed,
		making the formation of HMXBs via stellar scatterings more favorable.
		
		We submit that the HMXB formation rates inferred from our simulations
		be taken as a conservative estimate, with
		possible additional contributions from channels
		other than stellar scattering.
		
		\subsection{X-ray output}
		\label{subsec:X-rayspectra}
		As discussed in \S\ref{sec:intro}, HMXBs are believed to be a major
		source of X-rays in the early universe. 
		Observations of nearby star-forming galaxies
		suggest that their X-ray luminosities (which are dominated by HMXBs)
		scale linearly with their star formation rate
		(\cite{Grimm2003}, \cite{Gilfanov2004}, \cite{Persic2004}, \cite{xraysource2}).
	 In the linear regime, the X-ray luminosity of
		the local universe is given by \cite{Mineo2012},
		\begin{equation}
			\label{eq29}
			L^{{\rm local}}_{2-10~{\rm keV}}=3\times 10^{39}\times \frac{\rm SFR}{\Msol \yr^{-1}}~[\erg~{\rm s}^{-1}],
		\end{equation}
		where SFR is the star formation rate.

		In the following, based on the results of our simulations, 
		we are going to quantitatively evaluate the relation between X-ray
		luminosity and SFR, and compare it with equation (\ref{eq29}). We can write
		the X-ray luminosity as
		\begin{align}
			\label{eq28}
			L_{2-10~{\rm keV}}&=L_{\rm Edd}\times f_{\rm Edd}\times f_{2-10~{\rm keV}}\nonumber\\
			& \qquad \times t_{{\rm acc}} \times f_{{\rm sur}}\times f_{{\rm esc}}\times F_{\rm HMXB} \times {\rm SFR}\,.
		\end{align}
		
		Below, we discuss each quantity in equation (\ref{eq28}).
		\begin{enumerate}
			
			\item $L_{\rm Edd}$, the Eddington luminosity, which scales with the
			typical mass of the CO engine $M_{\rm CO}$ as
			$1.3\times 10^{38} (M_{\rm CO}/\Msol)~\erg\s^{-1}$.
			Note that, for stars which leave behind a BH remnant
			($M\gtrsim 25M_\odot$), we use the stellar mass as a proxy for
			the BH mass for simplicity, due to the theoretical uncertainties in
			evaluating the mass loss due to winds and during the transition to a
			BH. For stars with
			$M\lesssim 25M_\odot$, we consider two limiting cases: one in
			which all the stars form BHs, and one in which all the stars form
			NSs, for which we assume a typical mass value of 1.4~$\Msol$. 
			However, this uncertainty plays a very small role for our results
			due to the fact that the number of HMXBs with a star of $M<
			25\Msol$ is a small fraction of the total. 
			
			\item 
			$f_{\rm Edd}$, the typical ratio of the total radiative power emitted by HMXBs
			(the bolometric luminosity) to $L_{\rm Edd}$.
			If the typical luminosity of a HMXB during an active phase is
			Eddington, then $f_{\rm Edd}$ is effectively the mean duty cycle.
			Other studies have adopted values ranging from $0.1$ to $0.5$
			(\cite{xraysource2}, \cite{Belczynski2008}, \cite{Salvaterra+12}). 
			we take as our fiducial value $f_{\rm Edd}=0.1$.

			\item
			
			$f_{2-10~{\rm keV}}$, the fraction of the bolometric luminosity that
			is emitted between $2$ and $10\keV$.  Observational estimates vary
			between $0.1$ and $0.8$ (\cite{Sipior2003}, \cite{Migliari2006}).
			
			\item
			$t_{{\rm acc}}$, the time that a massive binary spends as a HMXB,
			with the less massive star donating mass to the CO companion.
			If the two stars form simultaneously and form a compact binary before the
			more massive member dies to become a CO,
			then this is simply $t_{2}-t_{1}$, the difference in the lifetimes of the stars.

			\item $f_{{\rm sur}}$, the fraction of HMXB candidates identified
			in our simulations that actually survive to become HMXBs.
			This quantity accounts for possible disruptions of binaries,
			due to (a) the merger of the stars during
			main sequence and post-main sequence evolution (\cite{Power2009});
			(b) the more massive star getting kicked following a supernova explosion 
			(\cite{Repetto2012}, \cite{Janka2013});
			and (c) subsequent disruptions by stellar scatterings that were not 
			captured by our simulations.
			
			\item $f_{{\rm esc}}$, the fraction of X-rays that escape the galaxy.
			Unless the environment of the HMXBs are Compton-thick, which is
			unlikely for the low-mass galaxies of interest, we expect $f_{{\rm esc}} \simlt 1$.

			\item $F_{\rm HMXB}$, the number of HMXBs formed per stellar mass.
			This is the main output of our simulations.\footnote{Note that $F_{\rm HMXB}$ has units $\Msol^{-1}$; we use the capital letter
			to distinguish it from the dimensionless fractions represented by $f$.}
			Whereas previous theoretical works had arrived at this value
			by extrapolating the locally observed value with an assumed redshift evolution,
			or with free parameters, here we provide an estimate based
			on suites of $N$-body simulations whose initial conditions
			are motivated by cosmological simulations of Pop~III star formation.
			A higher mass fraction in BHs should result in more X-ray production
			per unit star formation, 
			and a more pronounced effect on the IGM kinetic temperature
			and greater thermal feedback on early galaxy formation and evolution
			(\cite{Ripamonti+08}, \cite{Tanaka2012}).
			
		\end{enumerate}

		In all of our simulations independent of the number of stars in the group and collision configurations, we find
		$F_{{\rm HMXB}}\sim10^{-3}$, varying by less than a factor of $4$
		between the lowest and the highest values (see Table \ref{tab:tab4}).

		Finally, we can write $L_{2-10~{\rm keV}}$ as follows :  
		
		\begin{align}
			\label{eq30}
			\frac{L_{2-10~{\rm keV}}}{{\rm SFR}}
			=0.33&\times 
			\frac{M_{\rm BH}}{\Msol}\times \frac{f_{\rm Edd}}{0.1}\times \frac{f_{2-10~{\rm keV}}}{0.1}
			\times \frac{t_{{\rm acc}}}{\rm Myr}\nonumber\\
			&
			\times \frac{f_{{\rm sur}}}{0.5}
			\times \frac{f_{{\rm esc}}}{0.5}\times \frac{F_{\rm HMXB}}{10^{-3}\Msol^{-1}}
			\nonumber\\
			&\times \frac{10^{39}\erg\s^{-1}}{\Msol \yr^{-1}}\,.
		\end{align}
		Based on our choices of the factors, $f_{\rm Edd}=0.1$, $f_{2-10~{\rm
				keV}}=0.1$, $f_{{\rm sur}}=0.5$ and $f_{{\rm esc}}=0.5$, we can
		estimate the normalized X-ray luminosities per SFR, $ L_{2-10 {\rm
				keV}}/{\rm SFR}$. 
		We report this quantity for each of our models in 
		Table~\ref{tab:tab4}. It varies from a minimum of 37 to a maximum of 450 among
		the studied scenarios.

		These $L_{\rm X}$-to-SFR ratios are $\sim 40-150$ higher
		than what is observed in the local Universe.  This result is
		qualitatively consistent with the findings of \cite{BasuZych+13}
		and \cite{Kaaret2014}, who find an increase in the $L_{\rm
			X}$-to-SFR ratios toward $z\simgt 4$.
		Our high $L_{\rm X}$-to-SFR values stem from the large mass of the
		HMXBc primary, and the relatively low mass of the secondary.
		The former leads to a higher Eddington luminosity compared
		to typical stellar-mass BHs ($\sim 3\Msol$) in the local Universe,
		and the latter results in long stellar lifetimes, which in turn leads
		to longer $t_{\rm acc}$.

		\subsection{Implications for the thermal history of the IGM and the 21cm radiation}
		\label{subsec:implicationtothermalhistory}
		
		A higher $L_{\rm X}$-to-SFR ratio implies that IGM heating will
		occur earlier than commonly thought.
		The thermal history of the IGM can be probed in the $21\cm$ line, which
		is observable in absorption (or in emission),
		depending on whether the spin temperature of the IGM
		is below (or above) the CMB temperature.
		
		If the IGM heats early, as suggested by our estimates of the X-ray
		emission of early galaxies, the 21~cm absorption line appears earlier,
		and the ``dip'' as a function of redshift caused by adiabatic cooling
		is not as deep and not as sharp as in the case of late heating as it
		would otherwise (see Figure~2 in \cite{Fialkov2014}).  Another
		consequence of early, intense heating is that the temperature of the
		IGM could become high enough, to suppress the formation of low-mass
		galaxies (\cite{Ripamonti+08}) and the growth of their nuclear BHs
		(\cite{Tanaka2012}).

		\begin{table*}
			\centering
			\setlength\extrarowheight{1pt}
		\scriptsize
			\begin{tabulary}{0.5\linewidth}{c | c c | c c c c c c}
				\hline
				& 5-body ($n_{6})$ & 5-body ($n_{4}$) & \textit{Sce.~1} & \textit{Sce.~2} & \textit{Sce.~3} & \textit{Sce.~4A} & \textit{Sce.~4B}\\
				\hline
				$M_{1}$[$\Msol$] ($\tau_{\rm{life},1}[\Myr]$)	  &		110 (2.5)					&	88 (2.9) &	110 (2.5)	& 		120 (2.4) & 110 (2.4) &  160 (2.0) & 110 (2.4)	\\
				$M_{2}$[$\Msol$] ($\tau_{\rm{life},2}[\rm{Myr}]$)	    &  	12 (17)				&  11 (18)	&		18 (11)& 45 (4.8)	& 28 (7.1)& 63 (3.7) & 63 (3.7) \\
				$t_{\rm{acc}}~(=\tau_{\rm{life},2}-\tau_{\rm{life},1})$[Myr]   &  	14	&	15  & 8.6 	& 2.5 &	4.7 &	1.7 & 1.3  \\
				$F_{\rm HMXB}$ [$10^{-4}\Msol^{-1}$]	 &   4.6	&   4.6		&  15 &	11 & 9.0 & 4.2 & 8.4\\
				$L_{2-10 {\rm keV}}/{\rm SFR}$ $[10^{39}\rm{ erg s}^{-1}\Msol^{-1}\yr$] 	&	220 & 200 & 450 & 110 & 160 & 37 & 40 \\
				$L_{2-10 {\rm keV}}/L^{\rm local}_{2-10 {\rm keV}}$	&	75 & 67&  150 & 36 & 52 & 12 &  13\\
				\hline
				\hline
				$\frac{\eta_{\rm{GRB,Pop III}}}{\eta_{\rm{GRB,Pop I/II}}}$ &	2.2 & 2.2 & 7.0 & 5.4 & 4.3 & 2.0  & 4.0\\
				\hline
			\end{tabulary}
			
			\caption{Summary of the results of X-ray luminosity and GRB
				efficiencies. In the table, second and third columns
				correspond to 5-body calculations and the rest columns (from
				\textit{Sce.~1} to \textit{Sce.~4B}) are the results from
				10-body calculations with the number density of
				$10^{6}\rm{cm}^{-3}$.According to the ratio $L_{2-10~{\rm
						keV}}/L^{\rm local}_{2-10~{\rm keV}}$, where $L^{\rm
					local}_{2-10~{\rm keV}}/{\rm SFR}=3\times10^{39}\erg\s^{-1}$, it is
				expected that the 2-10~keV X-ray luminosity is $\sim 10^{2}$
				times larger than the X-ray luminosity of the local
				universe. These larger values essentially result from the
				large mass of the star 1.  In the bottom row, we have
				listed the LGRB efficiency, derived from each set of
				simulations. The higher HMXB formation rates, lead to larger
				efficiencies than the estimated value for Pop~I/II stars,
				$\eta_{\rm{GRB,Pop I/II}}\simeq 4.2\times10^{-6}$ with
				beaming factor of $1/50$ (\cite{Bromm2006}). }
			\label{tab:tab4}
		\end{table*}
		
		\subsection{Implications for Gamma Ray Bursts from Pop~III stars}
		\label{subsec:implicationtoGRB}
		
		As discussed in \S\ref{sec:intro},
		the fraction of HMXBs at high redshifts has potential implications
		for the expected rates of LGRBs from Pop~III stars. Binary systems 
		more easily satisfy the collapsar model (\cite{MacFadyen1999}) with respect to single stars
		(\cite{Cantiello2007}). Binary stars can spin up the helium core of the progenitor star
		via tidal coupling and spin-orbit locking.
		Further, RLOF can strip the hydrogen envelope during a common-envelope
		phase without reducing the rotation of the helium core \cite{Bromm2006}.
		This is especially important for Pop~III stars,
		whose heavier hydrogen envelopes would be more difficult to shed in isolation.
		Therefore, compact binary systems, or HMXBs, constitute a
		promising channel to produce LGRBs from Pop~III stars.
		
		We can use our results for the formation rates of Pop~III HMXBs
		to estimate the fraction of LGRBs from Pop~III stars.
		\cite{Bromm2006} quantified the GRB
		formation efficiency as
		\begin{equation}
			\eta_{\rm{GRB}}\simeq\eta_{\rm{BH}}\,\eta_{\rm{bin}}\,\eta_{\rm{close}}\,
			\eta_{\rm{beaming}},
		\end{equation}
		where $\eta_{\rm{BH}}$ is the number of
		BH-forming stars resulting from a given total stellar mass,
		$\eta_{\rm{bin}}$ is the binary fraction and $\eta_{\rm{close}}$ is
		the fraction of sufficiently close binaries to undergo RLOF.
		For Pop~I/II stars they calculated $\eta_{\rm BH}\simeq 1/(700\Msol)$.
		Combining this value with adopted values for the other parameters---$\eta_{\rm bin}\sim 0.5$,
		$\eta_{\rm close}\sim 0.3$, and
		$\eta_{\rm{beaming}}\simeq ({1}/{50})-({1}/{500})$---yields $\eta_{\rm GRB,PopI/II} \sim
		4.2\times (10^{-6} - 10^{-7}) \Msol^{-1}$.

		Our work is a first attempt to fill the gap between simplified guesses stemming from the absence of detailed
		calculations for the fraction of close binaries and our theoretical knowledge.
		Starting with $F_{\rm HMXB}=\eta_{\rm{BH}}\,\eta_{\rm{bin}}\,\eta_{\rm{close}}$ and using the same value of $\eta_{\rm{beaming}}\simeq ({1}/{50})-({1}/{500})$ (\cite{Bromm2006}),
		we then infer
		$\eta_{\rm{GRB, Pop~III}}\simeq 4.8\times(10^{-6}\sim10^{-7})
		\Msol^{-1}$ for the interacting 5-star case, and $2.2\times
		(10^{-5}\sim10^{-6}) \Msol^{-1}$ for the (most favorable) 10-body
		scenario.
		
		Therefore, our results suggest that LGRB rates from Pop~III stars could be comparable to
		or somewhat higher than the rates from Pop~I/II stars. 
		However, even with these rates, the probability of detecting a GRB
		from a Pop~III star remains very low and rather uncertain.
		The probability depends on
		the mass function of Pop~III stars,
		the likelihood that a Pop~III star produces a GRB,
		the luminosity function of these bursts,
		and how efficiently such high-redshift GRBs can be identified.
		Accordingly, there is a wide range of published theoretical results
		based on different model assumptions
		(\cite{Mesinger2005}, \cite{Bromm2006}, \cite{Salvaterra2008}, \cite{Ma2015}).
         In general, a detection of a GRB from a Pop~III star will likely require
		a long time baseline, provided by a combination of the current \textit{Swift}
		satellite and proposed future missions such as \textit{SVOM} (\cite{Schanne2010}) and 
		\textit{EXIST} (\cite{Grindlay2009}).

		\section{Summary}\label{sec:summary}
		
		In this study, we used $N$-body simulations of the first stars
		to explore the formation, evolution, disruption and energy output
		of Pop~III HMXBs.
		The code considers gravitational scattering of stars,
		dynamical friction, and the gravitational potential of ambient gas.
		
		The initial conditions for the simulations
		(i.e. IMF, typical star separation in the host haloes, ambient
		densities) are taken from two different sets of cosmological
		simulations of Pop~III formation, namely by \cite{stacy13} 
	       and \cite{Greif12}. We investigated star evolution in two backgound gas
		densities, a high-density case ($10^6 {\rm cm}^{-3}$), and a
		lower-density one ($10^4 {\rm cm}^{-3}$).

		Based on the handful of protostars per halo that are found in the
		works quoted above, we simulated systems with 5 stars and systems with
		10 stars. We found:
		\begin{enumerate}
			\item \textit{5-body simulations}:
		If stars form in compact groups separated by $\sim 10\AU$,
			as is expected from turbulent fragmentation,
			stellar scatterings lead to a significant HMXB formation rate.
			In particular, we found that HMXBs form at a rate of a few per $10^{4}\Msol$
			of stars formed, independent of the ambient gas density.
			\item \textit{10-body  simulations}:
			We simulated 10 stars on separations of $\sim 10\AU$,
			and evolved them as isolated quasi-Keplerian disks,
			or as two colliding groups with 5 stars each.
			For the latter, we ran several different sets of collision geometries.
			We found that the HMXB formation rate was a factor $\sim 1-3$ times 
			higher than for the $5$-body simulations, mainly due to the fact that
			the larger number of stars allowed for more hardening via stellar scattering.
		\end{enumerate}
		
		All of the simulations suggest an X-ray luminosity per
		unit star formation that is a factor $\sim 10^{2}$ higher than what is
		observed in the local Universe (under the assumption that other
		variables such as the X-ray escape fraction from galaxies and the duty
		cycle of HMXBs do not differ significantly).  These results are mostly
		due to \textit{the large mass of the most massive star of the HMXBc compared
			to that of the companion star, implying both a large $t_{\rm acc}$ as
			well as a higher luminosity of the remnant BH}.  The fact that we found
		little variation in this quantity across all of our simulations
		suggests that this is a robust estimate.  Additional factors, such as
		in-disk migration of nascent stars, could further increase the HMXB
		formation efficiency.

		A direct consequence is that X-rays can heat the IGM rapidly at Cosmic Dawn. Signals of early heating can be probed via
		the 21~cm line radiation. In addition to the
		implications for the thermal history of the IGM, these high formation rates
		of HMXBs per stellar mass imply a higher GRB formation efficiency from
		Pop~III stars in binaries. These predictions may be tested with a long baseline of
		observational data from {\em Swift} in combination with 
		future missions such as {\em SVOM, EXIST}.

\end{document}